\documentclass[useAMS,usenatbib]{mn2e}
\usepackage[english]{babel}
\usepackage{amsfonts}
\usepackage{amsmath}
\usepackage{amssymb}
\usepackage{multicol}
\usepackage{layout}
\usepackage{graphicx}

\usepackage{subfigure}

\usepackage{times}
\usepackage{natbib}

\newif\ifAMStwofonts
\AMStwofontstrue

\newcommand{\aj}{AJ}
\newcommand{\apj}{ApJ}
\newcommand{\apjs}{ApJS}
\newcommand{\aap}{A\&A}
\newcommand{\mnras}{MNRAS}
\newcommand{\prd}{Phys. Rev. D}
\newcommand{\jcap}{JCAP}

\graphicspath{{./fig/}}

\title[Spherical collapse model with shear and angular momentum in dark energy cosmologies]
    {Spherical collapse model with shear and angular momentum in dark energy cosmologies}
    
\author[A. Del Popolo et al.]
 {A. Del Popolo$^{1,2}$\thanks{adelpopolo@astro.iag.usp.br}, F. Pace$^{3}$, J. A. S. Lima$^{2}$\\
  $^{1}$ Astronomy Department, University of Catania, Italy\\
  $^{2}$ Departamento de Astronomia, Universidade de S\~ao Paulo, Rua do Mat\~ao 1226, 05508-900, S\~ao Paulo, SP,
Brazil\\
  $^{3}$ Institute of Cosmology and Gravitation, University of Portsmouth, Dennis Sciama Building, Portsmouth, PO1 3FX,
U.K.}

\date{Received \today; accepted ?}

\pagerange{\pageref{firstpage}--\pageref{lastpage}} \pubyear{2009}

\begin{document}
\label{firstpage}
\maketitle

\begin{abstract}
We study, for the first time, how shear and angular momentum modify typical parameters of the spherical collapse model,
in dark energy dominated universes. In particular, we study the linear density threshold for collapse
$\delta_\mathrm{c}$ and the virial overdensity $\Delta_\mathrm{V}$, for several dark-energy models and its influence
on the cumulative mass function. The equations of the spherical collapse are those obtained in \cite{Pace2010}, who 
used the fully nonlinear differential equation for the evolution of the density contrast derived from Newtonian
hydrodynamics, and assumed that dark energy is present only at the background level. With the introduction of the 
shear and rotation terms, the parameters of the spherical collapse model are now mass-dependant. The results of the
paper show, as expected, that the new terms considered in the spherical collapse model oppose the collapse of
perturbations on galactic scale giving rise to higher values of the linear overdensity parameter with respect to the
non-rotating case. We find a similar effect also for the virial overdensity parameter. For what concerns the mass
function, we find that its high mass tail is suppressed, while the low mass tail is slightly affected except in some
cases, e.g. the Chaplygin gas case.
\end{abstract}


\section{Introduction}
Till a decade ago, Universe was considered composed mainly by dark matter (DM) and characterized by a decelerating 
expansion. An important and surprising result coming from observational cosmology is the fact that high redshift
supernovae are less bright than expectations \citep{Riess1998,Perlmutter1999,Tonry2003}. This finding has been
interpreted as an acceleration in the expansion of the universe and that this acceleration is recent
\citep{Riess1998,Perlmutter1999,Knop2003,Riess2004,Astier2006}. This result has been confirmed by independent
observations: the baryon acoustic oscillation \citep{Tegmark2004a,Eisenstein2005,Percival2010}, the galaxy-galaxy
correlation function, giving important infos on the spatial distribution of large-scale structure
\citep{Tegmark2004b,Cole2005}, the angular spectrum of the CMBR temperature fluctuations 
\citep{Komatsu2011,Larson2011}, the Integrated Sachs-Wolfe (ISW) effect \citep{Ho2008}, globular clusters
\citep{Krauss2003, Dotter2011}, old high redshift galaxies (Alcaniz et al. 2003, 2005), and  galaxy clusters
\citep{Haiman2001,Allen2004,Allen2008,BasilakosPlionisLima2010} and weak lensing
\citep{Hoekstra2006,Jarvis2006}. The quoted accelerated expansion cannot be obtained in universes containing just 
matter homogeneously and isotropically distributed, while it can be obtained if the low-$z$ universe is filled in with 
a fluid with negative pressure, the so called {\it dark energy} (DE), with equation-of state-parameter $w<-1/3$. It is
possible to have accelerated universes without dark energy if one discards the homogeneity hypothesis on large scales
(e.g., Lemaitre-Tolman-Bondi (LTB) universes), using the back-reaction approach to dark energy \citep{Kolb2006}, 
gravitationally induced particle creation (Lima et al. 2009, 2010) or even modified models of gravity, like the $f(R)$
models \citep{Amendola2007}, $f(T)$ models \citep{Bengochea2009}, or the brane models \citep{Deffayet2001}.

The nature of dark energy is not understood to date, and this explains why in the past decade a large number of 
models for the origin and time evolution of DE have been proposed. In the $\Lambda$CDM model the DE is connected to the
energy of vacuum (cosmological constant) and the equation of state of DE, in this case, is simply $w=-1$. An extension 
of this model is obtained considering a scalar field with no or weak interaction with the matter component 
(quintessence models), and also phantom models, K-essence, or alternatively Chaplygin gas and Casimir effect.  

Cosmologists generally believe that structures in the universe grew via gravitational instability through the growth 
and collapse of primeval density perturbations originated in an inflationary phase of early Universe from quantum,
Gaussian distributed, fluctuations \citep{Guth1982,Hawking1982,Starobinsky1982,Bardeen1986}.

The presence of DE changes the rate of formation and growth of collapsed structures and large-scale structure, and 
consequently the distribution in size, in time and space of galaxies, quasars, supernovae, since they reside in 
collapsed structures. Moreover, DE, increasing the expansion rate, slows down the collapse of overdense structure and
its space-time fluctuations (if DE is not the cosmological constant) will give rise to DE haloes 
\citep{Creminelli2010} which will influence dark halos formation.

In principle, all the statistical information concerning the distribution of the dark matter and DE is contained in the 
probability distribution functions (PDFs) for the velocity, $\vec{v}$, and mass density fluctuation fields, $\delta$. 
The determination of the final moments of the quoted distribution from the starting ones require to know the exact 
dynamics ruling the evolution of the underlying field. Unfortunately the exact solution to the dynamical equations is 
known only in the linear regime of evolution of the quoted fields, and only approximate solutions are known for non-
linear stages. A popular analytical approach to study the non-linear evolution of perturbations of DM and DE is the
spherical collapse model (SCM) introduced in the seminal paper of \cite{Gunn1972} extended and improved in several  
following papers \citep{Fillmore1984,Bertschinger1985,Hoffman1985,Ryden1987,Avila-Reese1998,Subramanian2000,
Ascasibar2004,Williams2004}. Some papers \citep{Hoffman1986, Hoffman1989, Zaroubi1993} addressed the issue of the role 
of   velocity shear in the gravitational collapse.

The model describes how a spherical symmetric overdensity\footnote{A slightly overdense sphere, embedded in the 
universe, is a useful nonlinear model, as it behaves exactly as a closed sub-universe because of Birkhoff's theorem.}
decouples from the Hubble flow, slows down, turns around and collapse. In the model the overdensity is divided into 
bound mass, each one expanding with the Hubble flow from an initial comoving radius $x_i$ to a maximum one $x_m$
(usually named turn-around radius, $x_{ta}$), and then collapse. Non-linear processes convert the kinetic energy of 
collapse into random motions, giving rise to a ``virialized" structure.

The SCM proposed by \cite{Gunn1972} does not contain non-radial motions and angular momentum. The way to introduce 
angular momentum in the SCM, and its consequences, were studied in several papers
\citep{Ryden1987,Gurevich1988a,Gurevich1988b,White1992,Sikivie1997,Avila-Reese1998,Nusser2001,Hiotelis2002,  
LeDelliou2003,Ascasibar2004,Williams2004,Zukin2010}.

\cite{Ryden1987} were the first to relax the assumption of purely radial self-similar collapse by including
non-radial motions arising from secondary perturbations in the halo. \cite{Williams2004} used the same model
to show how angular momentum flattens the inner profile of haloes. The effect of non-radial motions on the mass profile
in a SCM was studied by \cite{Gurevich1988a,Gurevich1988b}.

\cite{White1992} applied a torque to particles in the shells during the initial expansion phase, and in order 
to preserve the spherical symmetry, assumed that the different particles acquire the same angular momentum but in 
independent randomly oriented directions. In \cite{Nusser2001}, particles acquire an angular momentum at turn around,
while before they move on radial orbits.  
Again, in order spherical symmetry is preserved and the angular momentum of each particle is conserved, particles have 
angular momenta distributed in random directions such that the mean angular momentum at any point in space is zero. 
Moreover, angular momentum is $\propto \sqrt{G M(<r_{\ast}) r_{\ast}}$\footnote{$r_{\ast}$ is the maximum radius of
oscillation of a particle} per unit mass \citep{White1992,Sikivie1997}, so no additional physical scale is introduced.
Other studies \citep{Hiotelis2002,LeDelliou2003,Ascasibar2004} introduced angular momentum in the SCM in a similar way
of the previous cited authors, and studied the structure of DM density profiles, reaching similar conclusions to that of
\cite{Williams2004}.

The SCM in the framework of dark energy cosmologies was studied by \cite{Mota2004,Abramo2007,Basilakos2010,Pace2010}. 
In particular
in \cite{Pace2010} were derived the equations for the SCM under the assumption that only DM can form clumps and that DE
is present as a background fluid \citep[see also][]{Fosalba1998a,Ohta2003,Ohta2004,Mota2004,Abramo2007}.   The
evolution of the overdensity $\delta$ is given by (Pace et al. 2010, 2012):
\begin{equation}\label{eqn:nleq}
 \begin{split}
  \ddot{\delta}+2H \dot{\delta}-\frac{4}{3}\frac{\dot{\delta}^2}{1+\delta}-
  4\pi G\bar{\rho} \delta(1+\delta)-\\
  (1+\delta)(\sigma^2-\omega^2) & = 0\;.
 \end{split}
\end{equation}
the shear term $\sigma^2=\sigma_{ij}\sigma^{ij}$ and the rotation term $\omega^2=\omega_{ij}\omega^{ij}$
are connected to the shear tensor, which is a symmetric traceless tensor, while the rotation is antisymmetric.
They are given by:
\begin{eqnarray}
 \sigma_{ij} & = & \frac{1}{2}\left(\frac{\partial u^j}{\partial x^i}+\frac{\partial u^i}{\partial x^j}\right)
 -\frac{1}{3}\theta\delta_{ij}\;, \\
 \omega_{ij} & = & \frac{1}{2}\left(\frac{\partial u^j}{\partial x^i}-\frac{\partial u^i}{\partial x^j}\right)\;.
\label{eq:tens}
\end{eqnarray}
where $\theta=\nabla_{\vec{x}}\cdot\vec{u}$ is the expansion.

Recalling that $\delta=\rho/\overline{\rho}-1=\left(a/R \right)^3-1$ (a is the scale factor 
and R the radius of the perturbation), and inserting it into Eq.~\ref{eqn:nleq}, it is easy to check that 
the evolution equation for $\delta$ reduces to the SCM \citep{Fosalba1998a,Engineer2000,Ohta2003}
\begin{equation}
\frac{d^2 R}{d t^2}= 4 \pi G \rho R -1/3  (\sigma^2-\omega^2) R= -\frac{GM}{R^2} -1/3 (\sigma^2-\omega^2) R  \;,
\end{equation}
comparable with the usual expression for the SCM with angular momentum \citep[e.g.][]{Peebles1993,Nusser2001,Zukin2010}:
\begin{equation}
\frac{d^2 R}{d t^2}= -\frac{GM}{R^2} +\frac{L^2}{M^2 R^3}= -\frac{GM}{R^2} + \frac{4}{25} \Omega^2 R,
\label{eqn:spher}
\end{equation}
where in the last expression we have used the momentum of inertia of a sphere, $I=2/5 M R^2$.

The previous argument shows that vorticity, $\omega$, is strictly connected to angular velocity, $\Omega$
(see also \cite{Chernin1993}, for a complete treatment of the interrelation of vorticity and angular momentum in
galaxies).

One assumption generally used when solving the SCM equations for the density contrast $\delta$ (Eq.~\ref{eqn:nleq})
is to neglect the shear, $\sigma$, and the rotation $\omega$. While the first assumption is correct, since for a sphere
the shear tensor vanishes, the rotation term, or angular momentum is not negligible. In fact, if we consider the ratio
of the rotational term and the gravitational one in Eq.~\ref{eqn:spher} we get $\frac{L^2}{M^3 R G}$ that for a spiral
galaxy like the Milky Way, with $L \simeq 2.5 \times 10^{74} g cm^2/s$ \citep{Ryden1987,Catelan1996a}, and radius 15 kpc
is of the order of 0.4, showing, as well known, that the rotation is not negligible in the case of galaxy sized 
perturbations. The quoted ratio is larger for smaller size perturbations (dwarf galaxies size perturbations) and smaller
for larger size perturbations (for clusters of galaxies the ratio is of the order of $10^{-6}$). The value of angular
momentum, $L$, or similarly $\Omega$, can be obtained and added to the SCM as described in
\cite{DelPopolo2009} or as described previously, assigning an angular momentum $\propto \sqrt{G M(<r_{\ast})r_{\ast}}$
at turn-around \citep[e.g.][]{White1992,Sikivie1997,Nusser2001}.

As previously stressed, the non-trivial role of angular momentum in the SCM has been pointed out in a noteworthy number 
of papers studying structure formation in DM dominated universes \citep[see
also][]{DelPopolo2009,Zukin2010,Cupani2011}. In a previous letter, \cite{DelPopolo2012} studied the effect of the term
$\sigma^2-\omega^2$ on the SCM parameters ($\delta_{\rm c}$ and $\Delta_{\rm V}$) for the Einstein-de Sitter (EdS) and
$\Lambda$CDM models, but it has never taken into consideration in the SCM in DE cosmologies.

In the present paper, we shall study how the typical parameters of the SCM (in Universes dominated by DE), namely 
the linear density threshold for collapse $\delta_\mathrm{c}$ and the virial overdensity $\Delta_\mathrm{V}$, are 
changed by a non-zero $\sigma$ and $\omega$ terms. In fact, any extension of the SCM should take into account the
effects of shear \citep{Engineer2000,DelPopolo2012} since shear induces contraction while vorticity induces expansion as
expected from a centrifugal effect.\\
We also study how angular momentum and shear influence the cumulative mass function.

The paper is organized as follows. In Sect.~\ref{sect:models}, we summarize the model used to obtain
$\delta_\mathrm{c}$ and the virial overdensity $\Delta_\mathrm{V}$. In Sect.~\ref{sect:results}, we describe the
results, and Sect.~\ref{sect:conclusions} is devoted to conclusions.

\section{Summary of the model}\label{sect:models}

The evolution equations of $\delta$ in the non-linear regime has been obtained and used in the framework of the 
spherical and ellipsoidal collapse, and structure formation by
\cite{Bernardeau1994,Padmanabhan1996,Ohta2003,Ohta2004,Abramo2007}. As a first step we assume that the fluid satisfies
the equation-of-state $P=w\rho c^2$. In addition, we also consider the Neo-newtonian expressions \citep{Lima1997} 
for the continuity, the
Euler equations, and the relativistic Poisson equation, namely:
\begin{eqnarray}
  \frac{\partial\rho}{\partial t}+\nabla_{\vec{r}}\cdot(\rho\vec{v})+
  \frac{P}{c^2}\nabla_{\vec{r}}\cdot\vec{v} & = & 0 \label{eqn:cnpert}\;,\\
  \frac{\partial\vec{v}}{\partial
    t}+(\vec{v}\cdot\nabla_{\vec{r}})\vec{v}+
  \nabla_{\vec{r}}\Phi & = & 0\;, \label{eqn:enpert}\\
  \nabla^2\Phi-4\pi G\left(\rho+\frac{3P}{c^2}\right) & = & 0\;,\label{eqn:pnpert}
\end{eqnarray}
where $\vec v$ is the velocity in three-space, $\Phi$ is the Newtonian gravitational potential and $\vec{r}$ is the 
physical coordinate.

The continuity equation for the mean background density can be written in the form 
\begin{equation}
 \dot{\bar{\rho}}+3H\left(\bar{\rho}+\frac{P}{c^2}\right)=0\;,
\end{equation}
where $\bar{\rho}=\frac{3H^2\Omega_{\mathrm{fluid}}}{8\pi G}$ is the background mass density of all contributions to 
the cosmic fluid, and $\Omega_{\mathrm{fluid}}$ is its density parameter.

Using comoving coordinates $\vec{x}=\vec{r}/a$, the perturbations equations can be written as
\begin{eqnarray}
 \dot{\delta}+(1+w)(1+\delta)\nabla_{\vec{x}}\cdot\vec{u} & = & 0\;, \label{eq:pertCont}\\
 \frac{\partial \vec{u}}{\partial t}+2H\vec{u}+(\vec{u}\cdot\nabla_{\vec{x}})\vec{u}+\frac{1}{a^2}\nabla_{\vec{x}}\phi &
 = & 0\;,\label{eq:pertEuler} \\
 \nabla_{\vec{x}}^2\phi-4\pi G(1+3w)a^2\bar{\rho}\delta & = & 0\;. \label{eq:pertPois}
\end{eqnarray}
where $H(a)$ is the Hubble function and $\vec{u}(\vec{x},t)$ is the comoving peculiar velocity.
Combining the previous equations, we get the non-linear evolution equation
\begin{equation}\label{eqn:nleqq}
 \begin{split}
  \ddot{\delta}+\left(2H-\frac{\dot{w}}{1+w}\right)\dot{\delta}-\frac{4+3w}{3(1+w)}\frac{\dot{\delta}^2}{1+\delta}-&\\
  4\pi G\bar{\rho}(1+w)(1+3w)\delta(1+\delta)-\\
  (1+w)(1+\delta)(\sigma^2-\omega^2) & = 0\;.
 \end{split}
\end{equation}
which is a generalization of Eq.~7 of \cite{Abramo2007} to the case of a non-spherical configuration of a rotating 
fluid.

In the case of dust ($w=0$), Eq.~\ref{eqn:nleqq} reads
\begin{equation}\label{eqn:nleq1}
\begin{split}
\ddot{\delta}+2H\dot{\delta}-\frac{4}{3}\frac{\dot{\delta}^2}{1+\delta}-
4\pi G\bar{\rho}\delta(1+\delta)-\\
(1+\delta)(\sigma^2-\omega^2) & = 0\;.
\end{split}
\end{equation}
which is Eq.~41 of \cite{Ohta2003}. 

In terms of the scale factor, $a$, the nonlinear equation driven the evolution of the overdensity contrast can be 
rewritten as:
\begin{equation}\label{eqn:wnldeq}
 \begin{split}
  \delta^{\prime\prime}+\left(\frac{3}{a}+\frac{E^\prime}{E} \right)
\delta^\prime-\frac{4}{3}\frac{\delta^{\prime 2}}{1+\delta}-
  \frac{3}{2}\frac{\Omega_{\mathrm{m},0}}{a^5 E^2(a)}\delta(1+\delta)-&\\
  \frac{1}{a^2H^2(a)}(1+\delta)(\sigma^2-\omega^2)&=0\;,
 \end{split}
\end{equation}
where $\Omega_m,0$ is the density parameter of the DM at $a=1$, $E(a)$ is given by
\begin{equation}\label{eq:e}
E(a)=\sqrt{\frac{\Omega_{\mathrm{m},0}}{a^3}+\frac{\Omega_{\mathrm{K},0}}{a^2}+ \Omega_{\mathrm{Q},0}g(a)}\;,
\end{equation}
where $g(a)$ is
\begin{equation}
g(a)=\exp{\left(-3\int_1^a\frac{1+w(a')}{a'}~da^{\prime}\right)}\;.
\end{equation}

Note that in Eq.~\ref{eqn:wnldeq} we corrected a typo present in Eq.~17 of \cite{Pace2010}.

In order to calculate the threshold for the collapse $\delta_{\rm c}$ and the virial overdensity, $\Delta_{\rm V}$, of
the SCM, we follow \cite{Pace2010}. We look for an initial density contrast such that the $\delta$ solving the
non-linear equation diverges at the chosen collapse time. Once the initial overdensity is found, we use this value as an
initial condition in the linearised equation
\begin{equation}\label{eqn:ldeq}
 \delta^{\prime\prime}+\left(\frac{3}{a}+\frac{E^\prime}{E}\right)\delta^\prime-\frac{3}{2}\frac{\Omega_{\mathrm{m},0}}{
a^5E^2}\delta=0\;,
\end{equation}
to get $\delta_\mathrm{c}$.

The initial conditions to solve the second-order differential equations are $\delta_{\rm i}$ (got as previously
described) and the initial rate of evolution, $\delta_\mathrm{i}^\prime$ is calculated as follows. We assume that at
early times the solution is a power law, therefore we can write $\delta_{\mathrm{i}}=Ba^n$, where $n$ is in general of
the order unity. The velocity is defined as the derivative with respect to the scale factor of the initial overdensity,
hence we can write it as $\delta_{\mathrm{i}}^{\prime}=n\delta_{\mathrm{i}}/a$.

The virial overdensity is obtained, as in \cite{Pace2010}, by using the definition
$\Delta_{\rm V}=\log{(\delta_{\rm nl}+1)}=\zeta(x/y)^3$, where $x=a/a_{\mathrm{ta}}$ is the normalised scale factor and
$y$ is the radius of the sphere normalised to its value at the turn-around.

The turn-around scale factor is obtained by solving Eq.~\ref{eqn:wnldeq} and determining the quantity
$\log(\delta_{\mathrm{nl}}+1)/a^3$. The virial overdensity at turn-around $\zeta$, is obtained by integrating
Eq.~\ref{eqn:wnldeq} up to $a_{\mathrm{ta}}$ and add the result to unity.

In order to integrate Eq. 15, we should explicit the $\sigma^2-\omega^2$ term.
The calculation of the term $\sigma^2-\omega^2$ is explained in detail in \cite{DelPopolo2012}. Here we simply 
summarize how we evaluated it. 

We first define with $\alpha$ the dimensionless ratio between the rotational and the gravitational term in 
Eq. (\ref{eqn:spher}):
\begin{equation}
 \alpha=\frac{L^2}{M^3 R G}\;,
\end{equation}
having different values for different scales, as already reported. 

We may calculate the same ratio between the gravitational and the extra term appearing in Eq. 15 thereby writing the
term extending the standard SCM as
\begin{equation}
 \frac{\sigma^2-\omega^2}{H_0^{2}}= -\frac{3}{2} \frac{\alpha\Omega_{\mathrm{m},0}}{a^3}\delta\;,
\end{equation}
as recently discussed in the literature \citep{DelPopolo2012}.\\
In order to obtain a value for $\delta_{\rm c}$  similar to the one obtained by \cite{Sheth1999}, we set $\alpha=0.05$
for galactic masses ($M\approx 10^{11}~M_{\odot}$) corresponding to a rotational velocity of $v_{\rm r}\approx 250~km/s$
and scaled it linearly towards higher masses and low velocities by assuming  a rotational velocity of nearly $10~km/s$ 
for galaxy cluster size objects ($M\approx 10^{15}~M_{\odot}$).

Given the above ansatz, the nonlinear equation that has to be satisfied by the overdensity contrast $\delta$ for a 
large class of noninteracting dark energy models reads:
\begin{equation}
 \delta^{\prime\prime}+\left(\frac{3}{a}+\frac{E^\prime}{E} \right)
 \delta^\prime-\frac{4}{3}\frac{\delta^{\prime 2}}{1+\delta}-
 \frac{3}{2}\frac{\Omega_{\mathrm{m},0}(1-\alpha)}{a^5E^2(a)}\delta(1+\delta)=0\;.
\label{eqn:lllll}
\end{equation}
It should be remarked that the DE models used are the same of
\cite{Pace2010}, namely the $\Lambda$CDM model, the quintessence models, phantom models and topological defects, and the
Chaplygin gas and Casimir effect. We refer to \cite{Pace2010} for more details on the models used.

\section{Results}\label{sect:results}

In this section we present results for the two main quantities derived in the framework of the SCM, in particular the
linear overdensity parameter $\delta_{\mathrm{c}}$ and the virial overdensity $\Delta_{\mathrm{V}}$. We assume as
reference model the $\Lambda$CDM model, with the following cosmological parameters: $\Omega_{\mathrm{m}}=0.274$,
$\Omega_{\mathrm{de}}=0.726$ and $h=0.7$.

\begin{figure}
\centering
\includegraphics[width=0.35\textwidth,angle=-90]{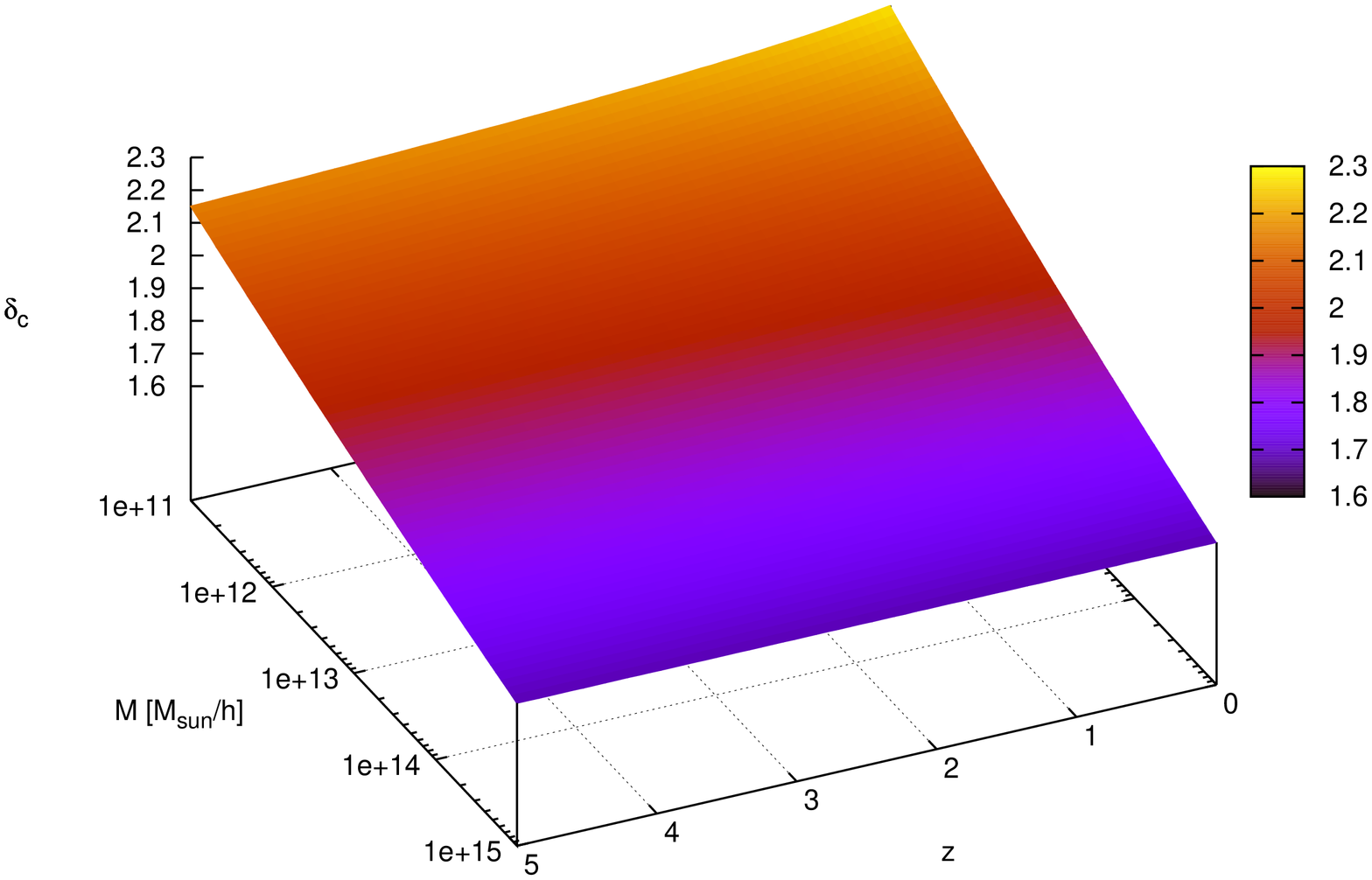}
\includegraphics[width=0.35\textwidth,angle=-90]{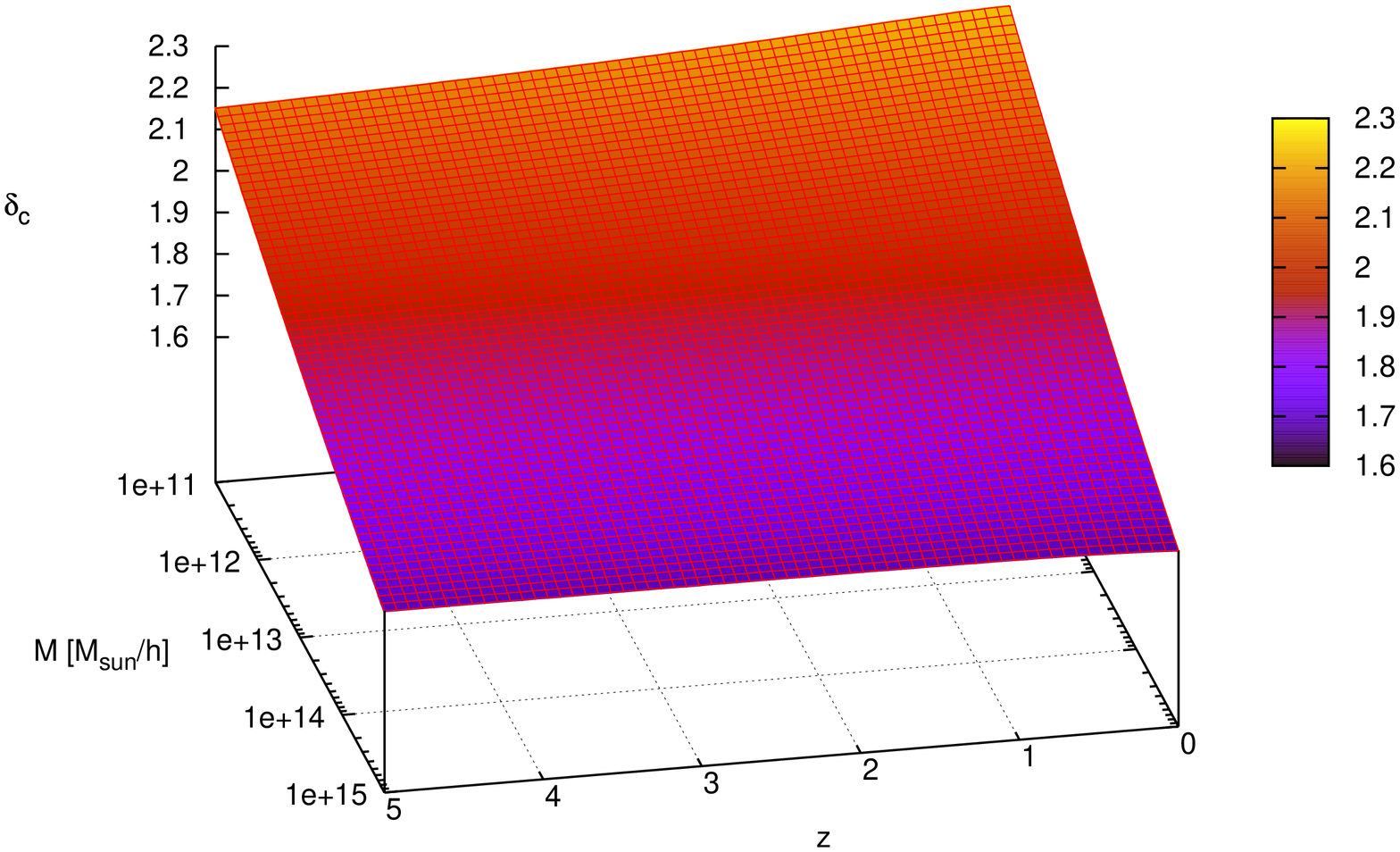}
\caption{Linear overdensity parameter $\delta_{\rm c}$ as a function of mass and redshift for the EdS model (upper
panel) and $\Lambda$CDM model (lower panel).}
\label{fig:delta3D}
\end{figure}

First of all, in Fig.~\ref{fig:delta3D}, we show the linear overdensity parameter $\delta_{\rm c}$ for the EdS model
(upper panel) and $\Lambda$CDM model (lower panel) as a function of the redshift and of the mass of the collapsing
object. Even if a projection of this plot was shown in \cite{DelPopolo2012}, we here propose the surface spanned by
$\delta_{\rm c}$ to briefly summarize how this quantity depends on the mass and on the redshift. 

An important result that it is worth to notice, is the time behaviour of $\delta_{\mathrm{c}}$. We observe that the
contribution of the term $\sigma^2-\omega^2$ is maximum at $z=0$ and it decreases with increasing redshift till
$\delta_{\mathrm{c}}$ reaches an approximately constant value, generally higher or equal to the value for the standard
SCM, according to the mass range considered. This is expected when we compare the nonlinear term with the gravitational
term. The net result is that of giving as source term a model with an effective matter density $(1-\alpha)$ times
smaller than the real matter density $\Omega_{\rm m}$ (only in the non-linear regime though). This can be interpreted as
an additional term counteracting the collapse even at high redshifts, making therefore $\delta_{\rm c}$ higher than the
standard value.

\begin{figure*}
 \centering{}
 \includegraphics[angle=-90,width=0.49\hsize]{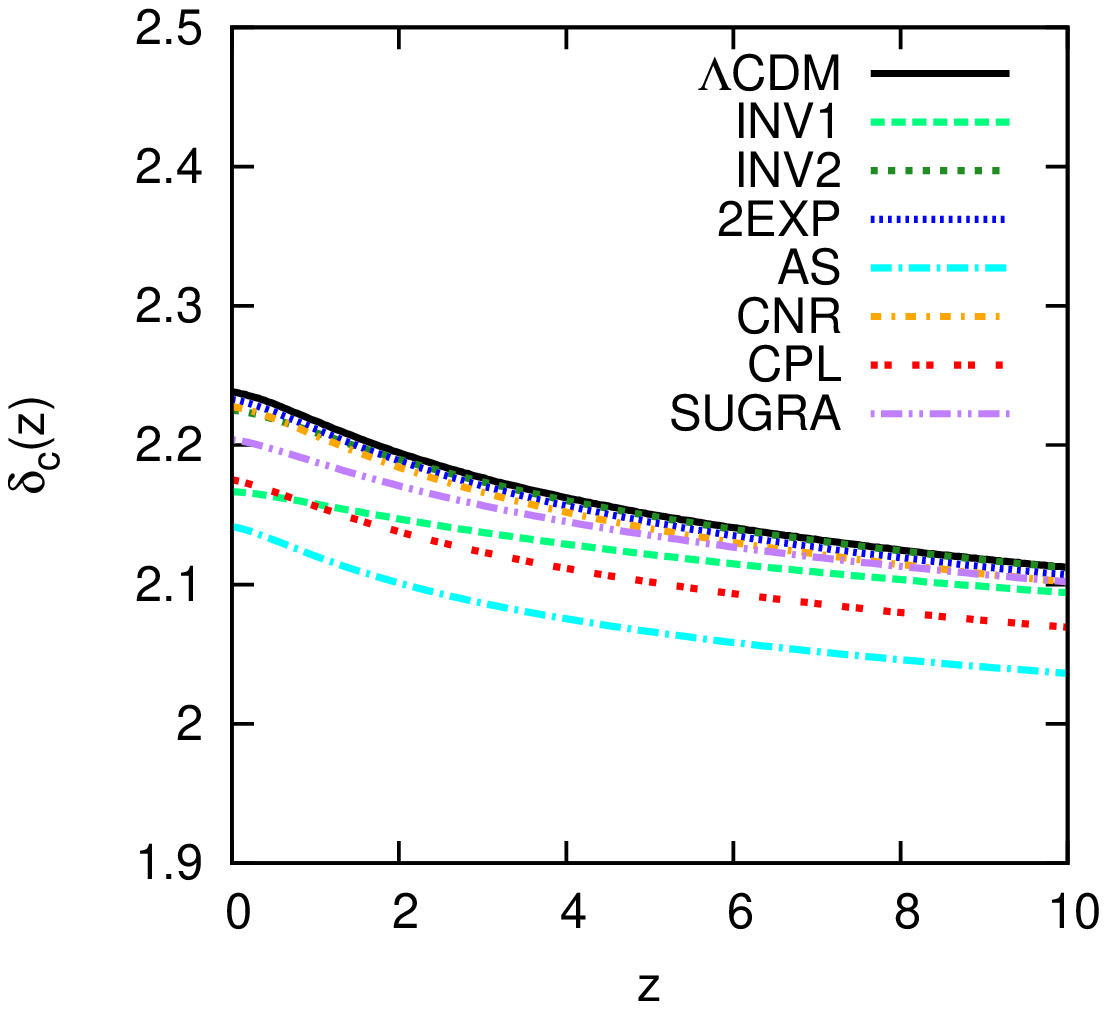}
 \includegraphics[angle=-90,width=0.49\hsize]{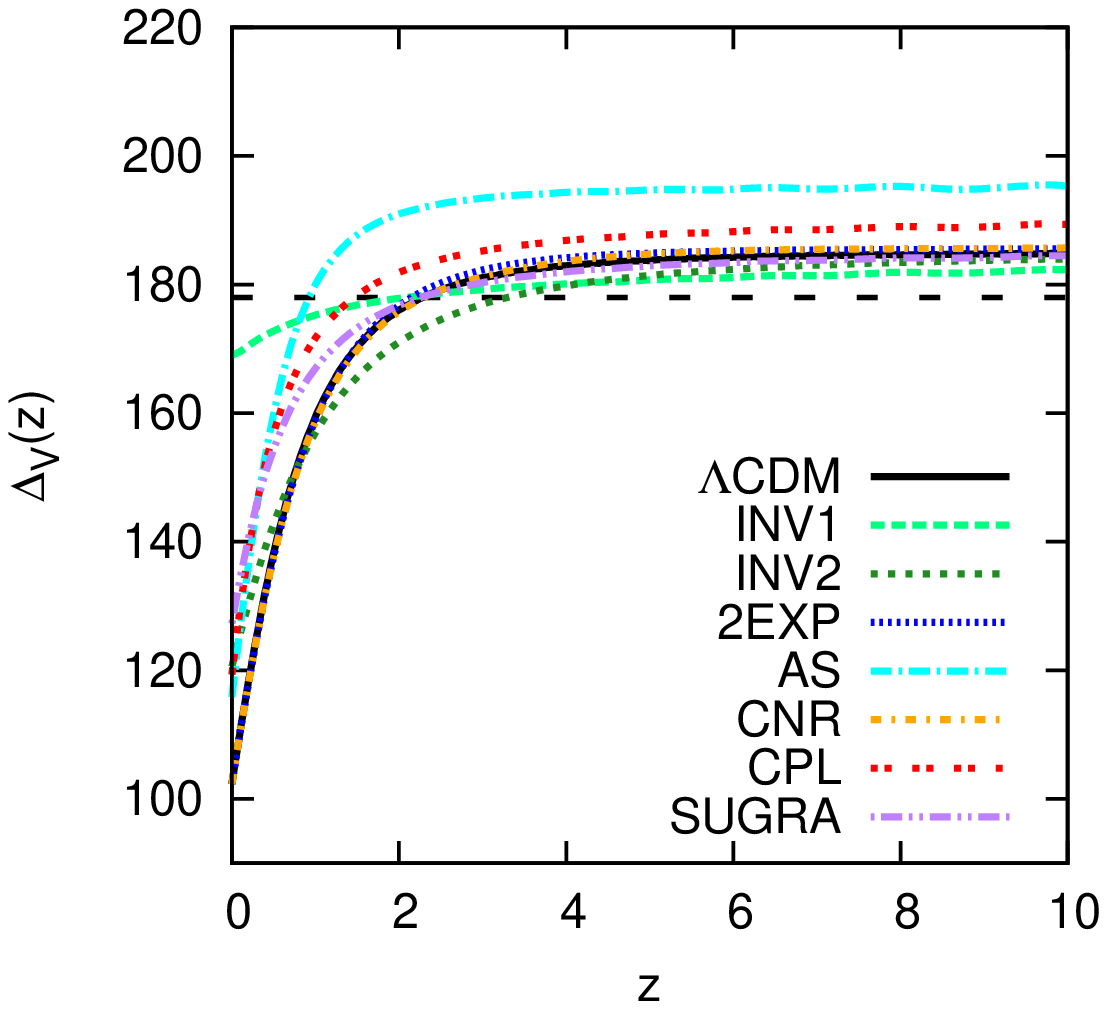}\hfill
 \includegraphics[angle=-90,width=0.49\hsize]{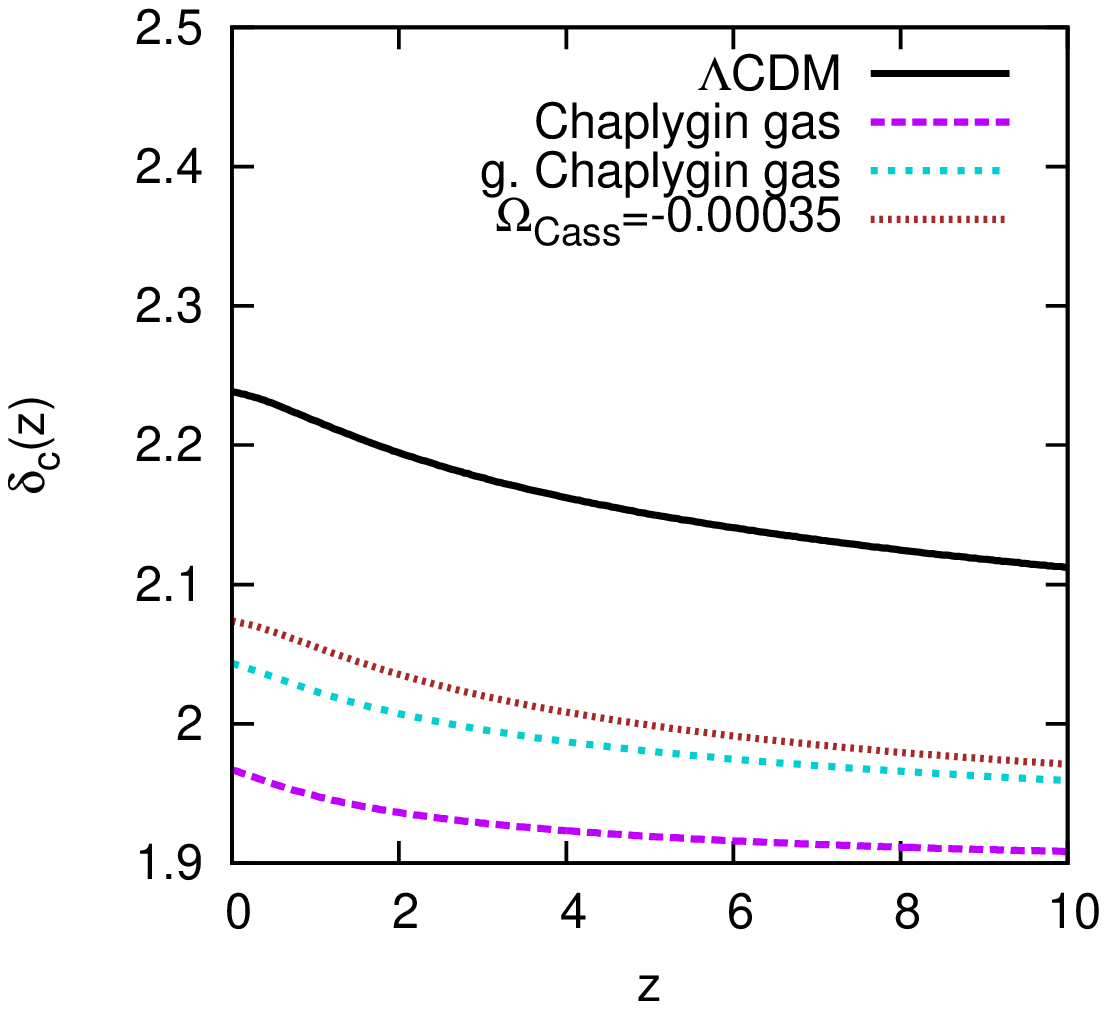}
 \includegraphics[angle=-90,width=0.49\hsize]{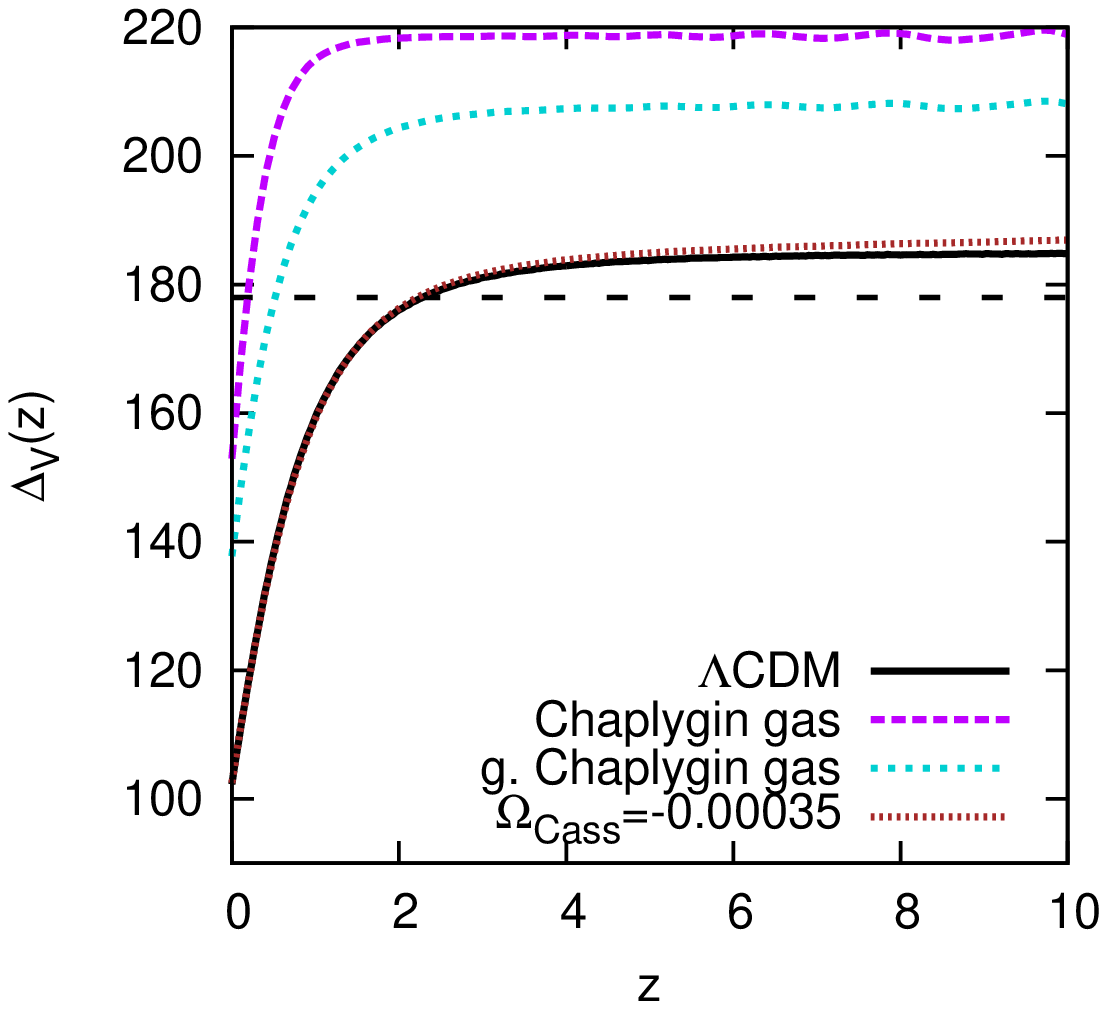}\hfill
 \includegraphics[angle=-90,width=0.49\hsize]{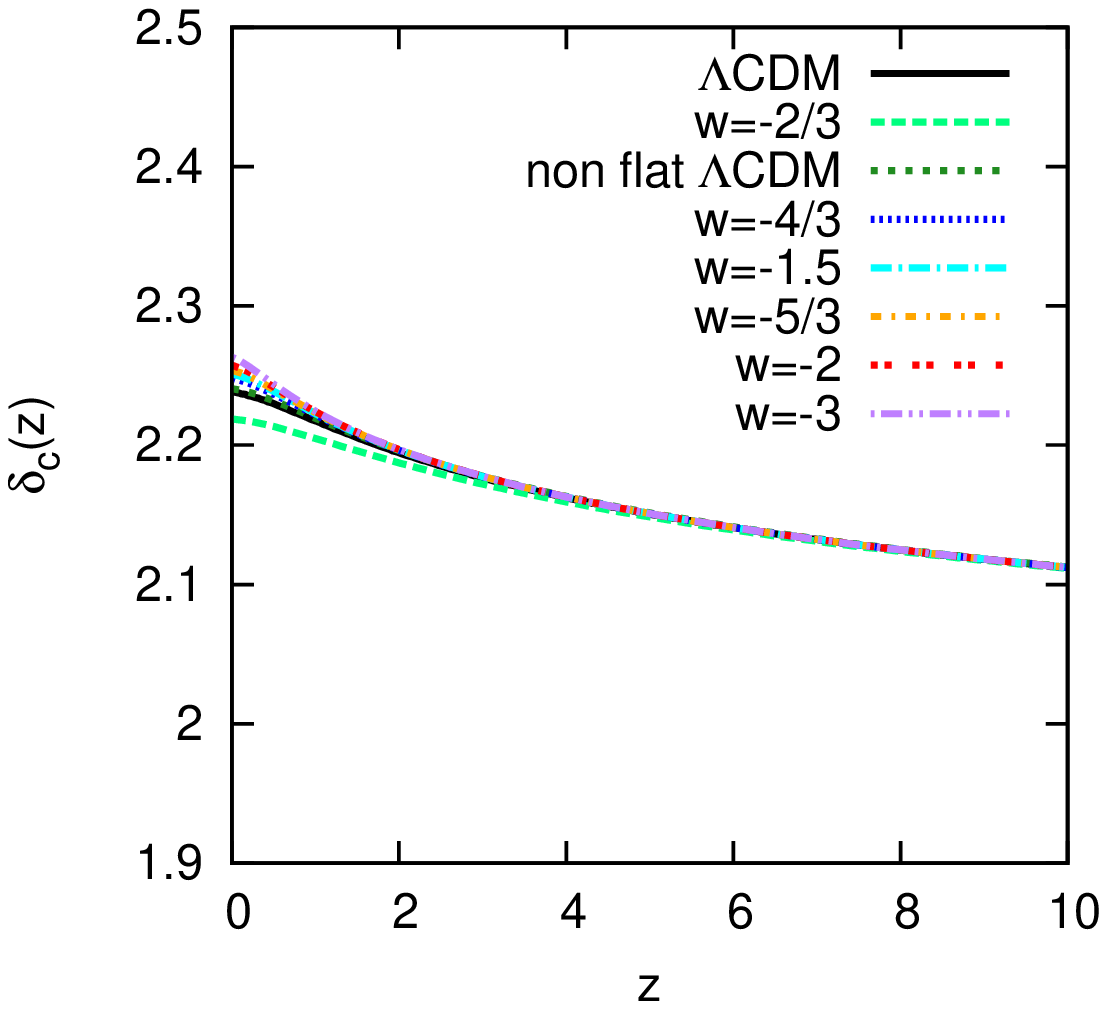}
 \includegraphics[angle=-90,width=0.49\hsize]{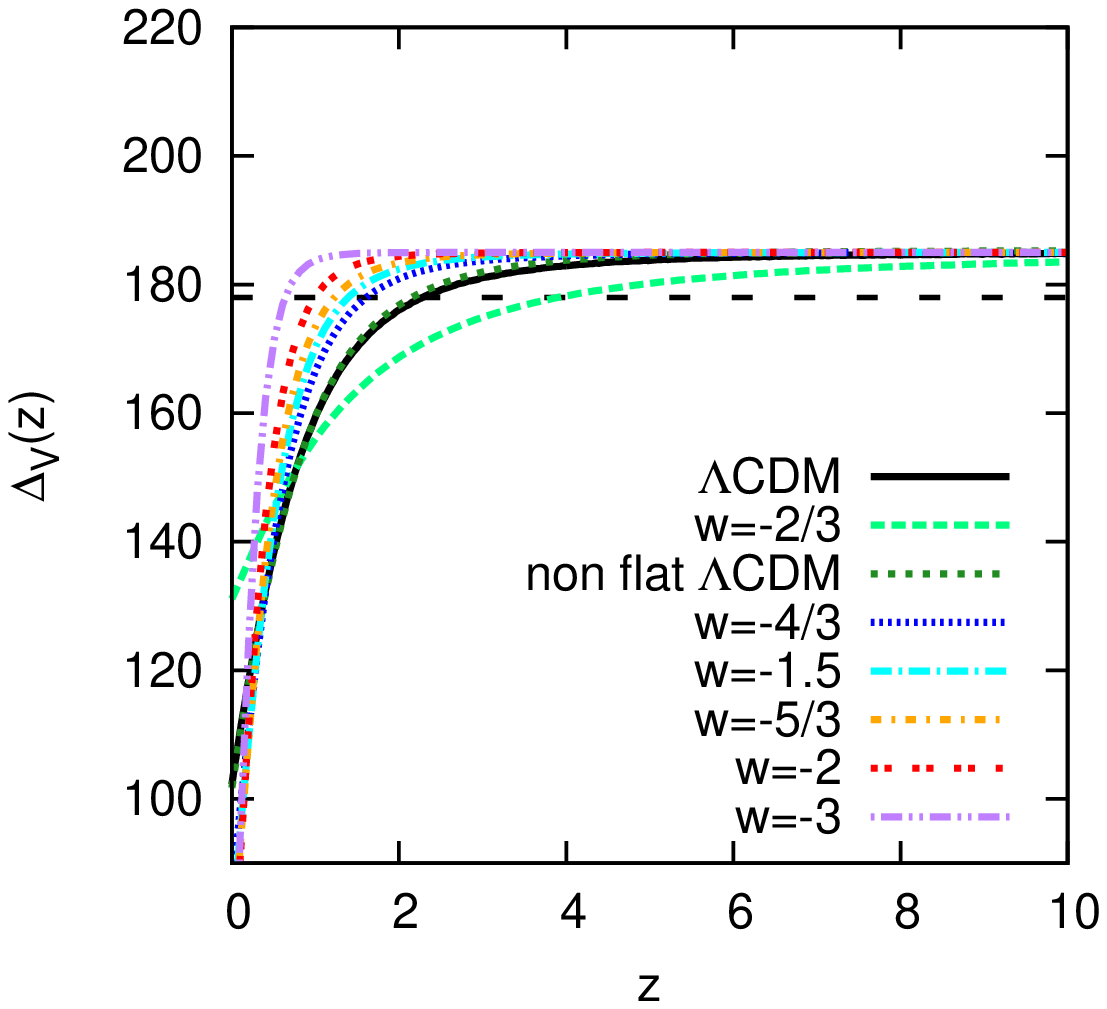}
 \caption{The left panels show the time evolution of the linear overdensity $\delta_{\mathrm{c}}(z)$, the right panels 
the time evolution for the virial overdensity $\Delta_{\mathrm{V}}(z)$ for the different classes of models. In all 
panels, the $\Lambda$CDM solution (black solid curve) is the reference model. All the curves assume a galactic mass for
the collapsing sphere. The upper panels present the quintessence models: the INV1 (INV2) model is shown with the
light-green dashed (dark-green short-dashed) curve, the 2EXP model with the blue dotted curve, the AS model with the
cyan dot-dashed curve, the CPL (CNR) model with the red dot-dotted (orange dot-short-dashed) curve and finally the SUGRA
model with the violet dot-dot-dashed curve. The middle panels show the Casimir effect (brown dotted curve) and the
(generalized) Chaplygin gas with the (turquoise short-dashed) magenta dashed curve. Finally the lower panels report the
solution for the models with constant equation-of-state parameter: the dark-green short-dashed curve stands for the
non-flat $\Lambda$CDM model, the light-green dashed curve for the model with $w=-2/3$, the blue dotted curve represents
the model with $w=-4/3$, the cyan dot-dashed curve the model with $w=-1.5$, the orange dot-short-dashed curve the model
with $w=-5/3$, the red dot-dotted curve the model with $w=-2$ and finally the violet dot-dot-dashed curve shows the
model with $w=-3$.}
 \label{fig:spc}
\end{figure*}

In Fig.~\ref{fig:spc} instead we present results for the different dark energy models considered in this work. On the
left panel we show results for the linear overdensity parameter $\delta_{\mathrm{c}}$ while on the right panel we show
the expected values for the virial overdensity $\Delta_{\mathrm{V}}$.\\
Because of the consideration expressed above regarding the EdS model, we will consider as reference the standard
$\Lambda$CDM model (with $\alpha=0$) and to maximize the effect of the non linear term, all the figures show results
for galactic masses ($M\approx 10^{11}~M_{\odot}/h$).

For the first class of models (INV1, INV2, 2EXP, AS, CNR, CPL, SUGRA) we see that the models are in general very similar
to each other and they slightly differ from each other, with differences at most of the order of $4.5\%$ for the AS
model. The INV1, CNR and 2EXP models are basically indistinguishable from the $\Lambda$CDM model. We interpret this
result as due to the equation of state of the models considered. As shown in \cite{Pace2010}, the INV1 model
has an equation of state basically constant over the whole cosmic history, but its present value is quite different from
all the other models, being $w_0\approx -0.4$, while for all the other models is $-1\leq w_0\leq -0.8$.\\
Comparing our present results with the ones of the upper left panel in Fig.~4 of \cite{Pace2010}, we see that the
behaviour of the models is very similar. The inclusion of the non-linear term just changes the values of the linear
overdensity parameter, but not the respective ratios with the $\Lambda$CDM model.

For the second group of models (Casimir and (generalized) Chaplygin gas) we obtain very different results from
\cite{Pace2010}. While there only the generalised Chaplygin gas was substantially different from the $\Lambda$CDM
model, now all the models here considered differ much from the reference model. This shows how the non-linear
additional term is very sensitive to the equation of state considered. Moreover, none of the models recovers the
extended $\Lambda$CDM model at high redshifts.

The bottom panel in the left column is devoted to the phantom models ($w<-1$) and to a non-flat $\Lambda$CDM model. All
the models present very similar results and small differences appear at small redshifts ($z\lesssim 1$), and for
redshifts $z\gtrsim 2$, all the models are identical. Models differing most at $z\approx 0$ are the models with
$w=-2/3$ and $w=-3$ having values for $\delta_{\rm c}$ respectively lower and higher than the $\Lambda$CDM model. This
is in agreement with results of \cite{Pace2010}, showing once again that a super-negative equation-of-state affects 
only slightly the structure formation process. In particular, the more negative it is, the higher is the linear
overdensity parameter. We also notice that a small amount of curvature, does not influence our results significantly.

Comparing these results, with the results from Fig.~4 in \cite{Pace2010}, we can appreciate the interplay between
the term $\sigma^2-\omega^2$ and a dynamical dark energy equation of state.\\
All the models studied with a time-varying dark energy equation-of-state parameter show that the collapse, even if
retarded by the inclusion of the shear and rotation, is easier as compared with the $\Lambda$CDM model. In this case,
with easier we mean that the values for the extended $\delta_{\mathrm{c}}(z)$ are smaller than for the reference model.
This is expected and it has the same explanation as for the usual case. Since at early times the amount of dark
energy is higher, we need structures to grow faster in order to observe cosmic structures today. This is plausible,
since the linear overdensity parameter represents the time evolution of the initial overdensity, whose evolution is
dictated by the growth rate that is described by the same differential equation.
In other words, since at early times the amount of dark energy is higher, we need lower values of $\delta_c$ to 
have objects collapsing. This is analogous to the case of the linear growth factor, since the equation to be solved is 
the same.\\
Opposite is instead the behaviour for the phantom models, in which case we notice that with a more negative equation of 
state, the collapse is retarded more severely. This is in agreement with Fig.~4 of \cite{Pace2010}, where
phantom models had an higher $\delta_{\mathrm{c}}$: since the expansion goes so fast, the collapse is strongly
suppressed and with respect to the reference models, higher and higher initial overdensities are required in order to
have collapsed objects today.

On the right panels, we show results for the quantity $\Delta_{\rm V}$. Also in this case we limit ourselves to
galactic masses. We notice that for the quintessence models in general the virial overdensity shows higher values than
for the extended $\Lambda$CDM model, except for the INV2 model. This is opposite to what found for the standard case,
where all the models had smaller values than the $\Lambda$CDM model. Also in this case none of the models approximates
the extended $\Lambda$CDM model at high redshifts. This is not the result of the additional non-linear term only, but
also of the influence of the dark energy equation of state, consistently with results from \cite{Pace2010}.

For the phantom models, results are very similar to the usual case. Virial overdensity parameter is higher than the
extended $\Lambda$CDM one if $w<-1$ and lower for the model with $w=-2/3$, in agreement with the linear overdensity
parameter. As shown in \cite{Pace2010}, at higher redshifts, all the phantom models reduce to the $\Lambda$CDM model.

Differences in the linear overdensity parameter reflect in the differential mass function. In this work we decide to
use the parametric form by \cite{Sheth1999,Sheth2001,Sheth2002}. We consider three different redshifts, namely $z=0,
0.5, 1$.

At this point, it is worth noticing that in the first paper of \cite{Sheth1999} the mass function was calculated 
as a fit to numerical simulations. Later on, \cite{Sheth2001} and \cite{Sheth2002} showed that the effects of
non-sphericity (shear and tides) introduce a mass dependence in the collapse threshold (see Eq. 4 in \cite{Sheth2001}
and following discussion). By using this threshold as the barrier in the excursion set approach one gets a mass function
in good agreement with simulations (see Eq. 5 and the discussion in the last part of Sect. 2.2. of \cite{Sheth2001}, and
moreover \cite{Sheth2002}).

For the $\Lambda$CDM model we choose as power spectrum normalization the value $\sigma_8=0.776$. Since we want that
perturbations at the CMB epoch are the same for all the models, we normalize the dark energy model according to the
formula

\begin{equation}
 \sigma_{8,\rm DE}=\sigma_{8,\Lambda\rm CDM}\frac{D_{+,\Lambda\rm CDM}(a_{\rm CMB})}{D_{+,\rm DE}(a_{\rm CMB})}\;,
\end{equation}
where $D_{+}(a_{\rm CMB})$ is the growth factor at the CMB epoch.

In Table~\ref{tab:sigma}, we display the different normalizations used for the models considered in this work ($K$
is the curvature parameter).

\begin{table}
 \caption{Table with the power spectrum normalization for the different dark energy models ($K$ is the
curvature parameter).}
 \label{tab:sigma}
 \begin{center}
  \begin{tabular}{c|c}
   \hline
   \hline
   Model & $\sigma_8$\\
   \hline
   $\Lambda$CDM, K$=0$ & 0.776 \\
   $\Lambda$CDM, K$\neq 0$ & 0.793 \\
   INV1 & 0.428 \\
   INV2 & 0.707 \\
   2EXP & 0.739 \\
   AS & 0.319 \\
   CNR & 0.732 \\
   CPL & 0.444 \\
   SUGRA & 0.578 \\
   Chaplygin gas & 0.066 \\
   g. Chaplygin gas & 0.133 \\
   Casimir & 0.420 \\
   Phantom ($w=-2/3$) & 0.674 \\
   Phantom ($w=-4/3$) & 0.834 \\
   Phantom ($w=-1.5$) & 0.854 \\
   Phantom ($w=-5/3$) & 0.870 \\
   Phantom ($w=-2$) & 0.894 \\
   Phantom ($w=-3$) & 0.936 \\
   \hline
   \hline
  \end{tabular}
 \end{center}
\end{table}

\begin{figure}
 \centering{}
 \includegraphics[width=0.7\hsize,angle=-90]{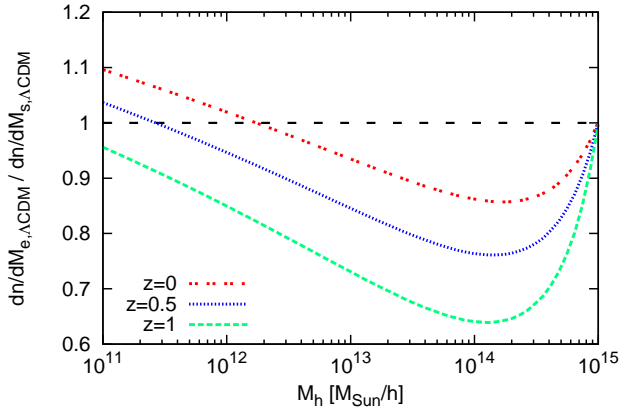}
 \caption{Ratio between the differential mass function of the extended and standard $\Lambda$CDM model. The curves
represent three different redshifts: $z=0$ (dotted red curve), $z=0.5$ (short-dashed blue curve), $z=1$ (dashed green
curve).}
 \label{fig:mfLCDM}
\end{figure}

In Fig.~\ref{fig:mfLCDM} we compare the differential mass function for the $\Lambda$CDM model in the standard and in
the extended SCM. \\
Analysing the three curves, we can appreciate the effect of the term $\sigma^2-\omega^2$ on the mass function. On small
masses, the mass function is largely independent of the cosmological model, but it depends strongly on $\delta_{\rm
c}$. Since in the extended SCM this is higher, we observe an increase in the number of objects at galactic scale at
$z=0$, to decrease at higher redshifts where the contribution of the non-linear term decreases. We observe a general
decrement in the number of objects at high masses (up to $M\approx 10^{14}~M_{\odot}/h$) to increase again to unity
for masses of the order of $10^{15}~M_{\odot}/h$. This is explained by the fact that at such masses, the linear
overdensity parameter are practically the same, therefore the mass function must not change.

In this concern, it should be recalled that shear and rotation have the maximum effect on $\delta_c$, at galactic scale
{(see Fig.~\ref{fig:delta3D})}. However, in the calculation of the ratio between the differential mass functions
(Fig.~\ref{fig:mfLCDM}), beside $\delta_c$ we need to take into account the factor $\sigma(M)$, the r.m.s. of mass
overdensity. Now, recalling the Sheth \& Tormen multiplicity function
\begin{equation}
f_{ST}=A \sqrt{ \frac{2a}{\pi}} \left[ 1+ \left(\frac{\sigma(M)^2}{\delta_c^2 a}\right)^p \right ]
\frac{\delta_c}{\sigma(M)} e^{-\frac{a\delta_c^2}{2 \sigma(M)^2}}\;,
\label{eq:ST}
\end{equation}
with $A=0.3222$, $a=0.707$, and $p=0.3$, we have that at galactic scale the dominant term in Eq.~\ref{eq:ST} is the 
term $\frac{\delta_c}{\sigma}$, consequently the ratio $f_{\rm ST,extended}/f_{\rm ST,\Lambda CDM} \simeq
\delta_{\rm c,extended}/\delta_{\rm c,\Lambda CDM}$. At larger masses, the effect of rotation and shear diminishes with
the consequence that $\delta_{\rm c,extended} \simeq \delta_{\rm c,\Lambda CDM}$ and that the $\sigma(M)$ term is fixing
the value of the $f_{\rm ST,extended}/f_{\rm ST,\Lambda CDM}$ ratio.

{At this point it is necessary a deeper discussion of the results shown in Fig.~\ref{fig:mfLCDM}. The ST mass
function generalises the Press \& Schechter (PS) \citep{Press1974} mass function to include the effects of shear and
tidal forces with respect to the simpler spherical collapse model. In doing so it is necessary to consider the
ellipsoidal collapse model and the corresponding linear overdensity parameter $\delta_{\rm ec}$. Differently from the
spherical collapse model, now $\delta_{\rm ec}$ is not only a function of time, but of mass too and the relation between
$\delta_{\rm ec}$ and $\delta_{\rm c}$ ($\delta_{\rm sc}$ in \cite{Sheth2001}) is given by Eq.~4 in \cite{Sheth2001}.
The moving barrier for the random walk is now set equal to $\delta_{\rm ec}(M,z)$ and a good approximation to it is
then given by their Eq.~5. The ST mass function fits quite well the results of N-body simulations, but as shown in
Fig.~\ref{fig:mfLCDM}, for masses $M\approx 10^{14}M_{\odot}/h$ our predictions, including rotation on top of the
ellipsoidal collapse, predict approximately $40\%$ less objects at $z=1$ than the standard $\Lambda$CDM model. This
would be easily checked with a big enough spanned volume. Moreover one could identify ellipticity and rotation if halos
acquire angular momentum by misalignment with the surrounding tidal field. Following this line of thought one could
further use the $\delta_{\rm c}(M,z)$ predicted by the extended spherical collapse model as the correct moving barrier.
Therefore by direct comparison of expressions (2), (4) and (5) in \cite{Sheth2001} we substitute 
$\delta_{\rm c,extended}$ directly into their expression (2) (equivalent to the PS multiplicity function). Results are
shown in Fig.~\ref{fig:mulfunc} for the three redshifts of Fig.~\ref{fig:mfLCDM}.

\begin{figure}
 \centering{}
 \includegraphics[width=0.7\hsize,angle=-90]{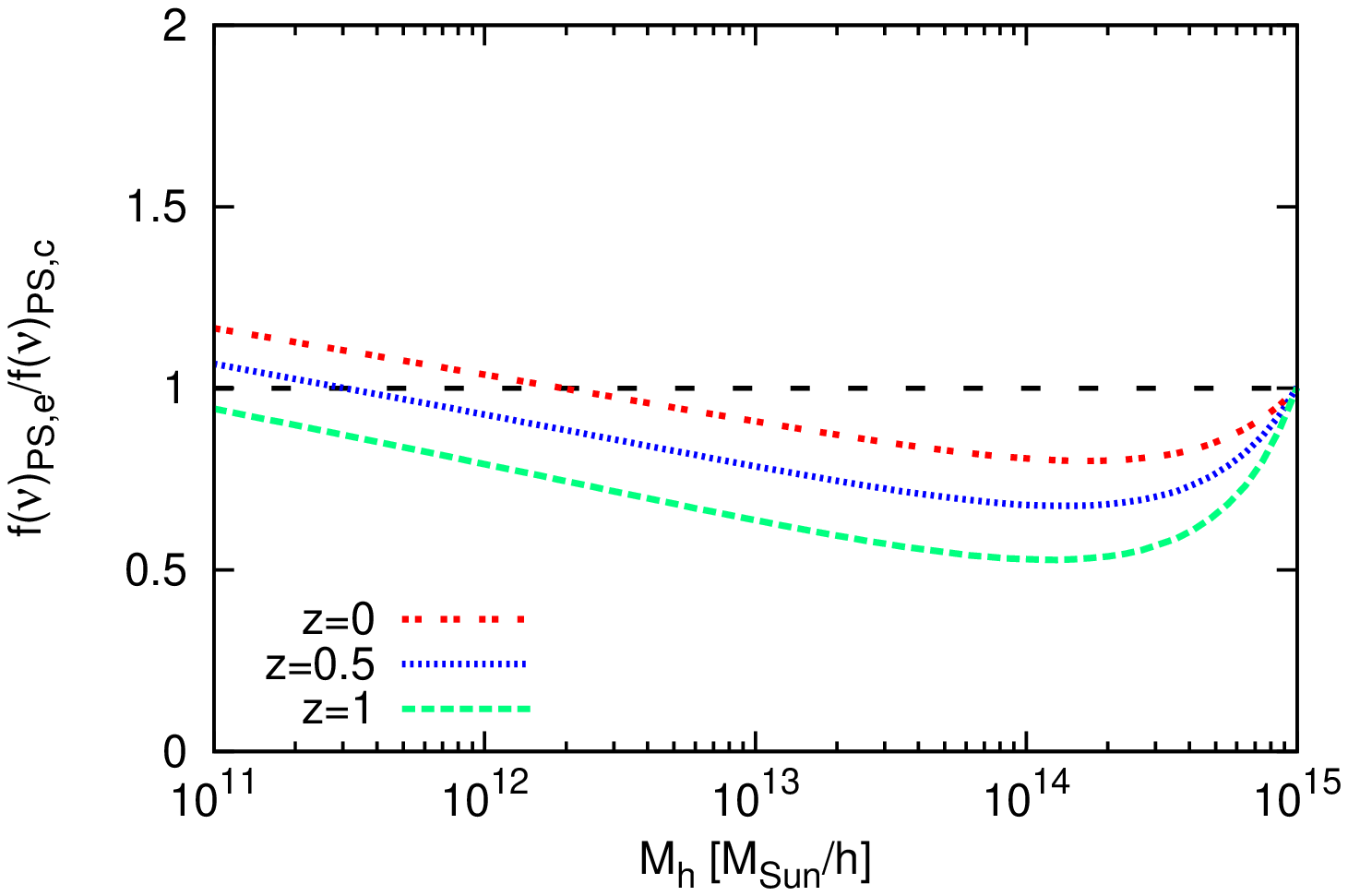}
 \includegraphics[width=0.7\hsize,angle=-90]{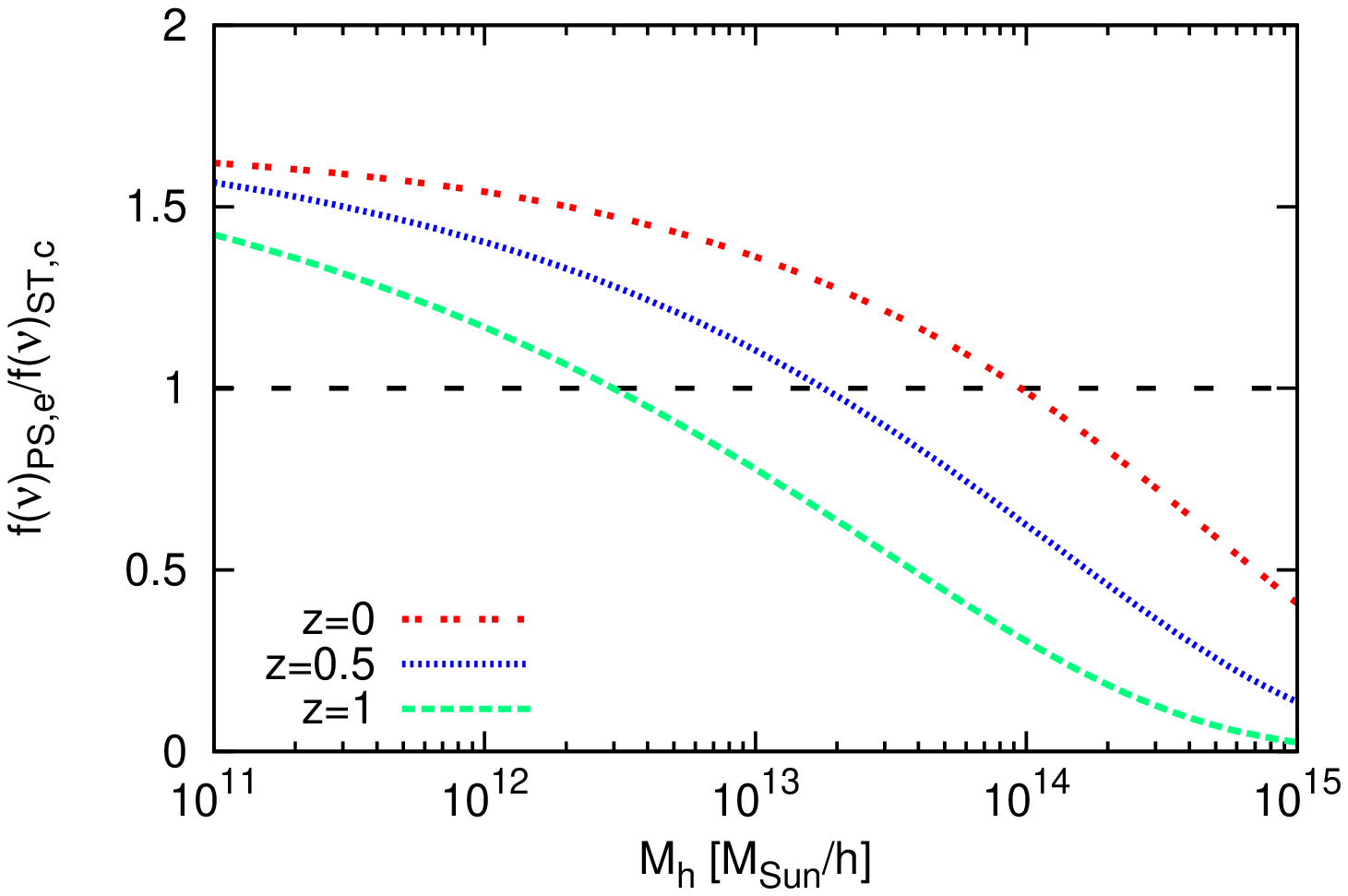}
 \caption{Ratio between the PS multiplicity function evaluated with $\delta_{\rm c,extended}$ and $\delta_{\rm c,\Lambda
CDM}$ (upper panel) and ratio between the PS multiplicity function with $\delta_{\rm c,extended}$ and the ST
 multiplicity function with $\delta_{\rm c,\Lambda CDM}$. Line styles and colours are as in Fig.~\ref{fig:mfLCDM}.}
\label{fig:mulfunc}
\end{figure}

In the upper panel we show the ratio of the PS multiplicity function evaluated with the $\delta_{\rm c,extended}$ and
$\delta_{\rm c,\Lambda CDM}$ linear overdensity parameters while in the lower panel we compute the ratio of the PS
multiplicity function evaluated with the $\delta_{\rm c,extended}$ linear overdensity parameter with the ST
multiplicity function evaluated with the $\delta_{\rm c,\Lambda CDM}$ linear overdensity parameter. As it is evident,
when we compute the ratio using the same multiplicity function, we obtain a very similar behaviour as in
Fig.~\ref{fig:mfLCDM}, even if quantitatively slightly different. When we instead identify rotation with ellipticity
and use directly the PS multiplicity function and compare it with the ST multiplicity function, we observe a totally
different behavior. This can be explained not only in terms of different overdensity parameters, but also remembering
that the PS mass function predicts more (less) objects at low (high) masses with respect to the ST parametrization. Our
results also show that if we want to reproduce the results of the ST mass function using the extended spherical
collapse model, we need to modify the moving barrier and find a new parametrization for the multiplicity function. This
goes beyond the purpose of this work, therefore, taking into account these caveats we will assume the ST mass function
as the correct one.}

\begin{figure*}
 \centering{}$
 \begin{array}{ccc}
 \includegraphics[width=0.22\textwidth,angle=-90]{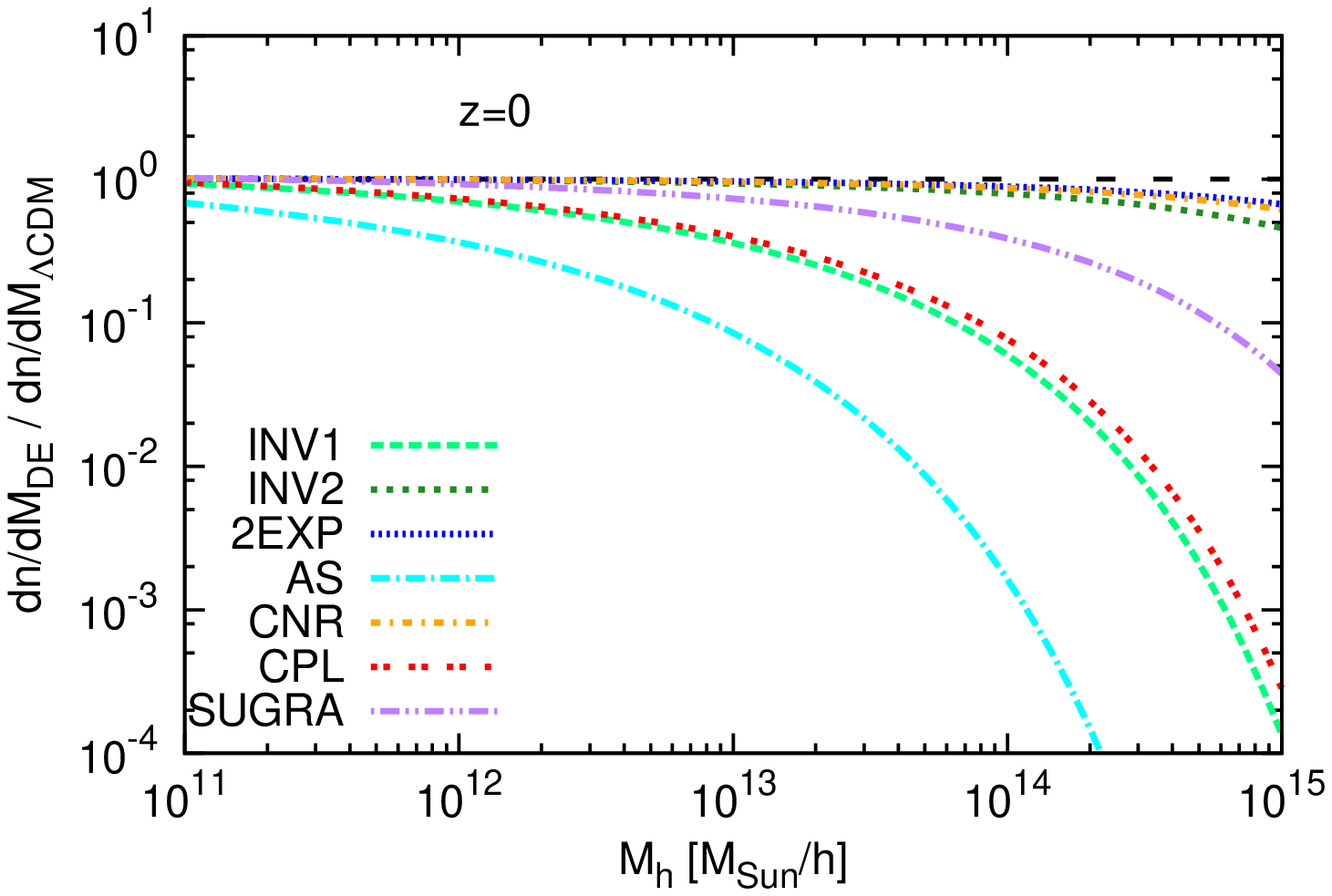} &
 \includegraphics[width=0.22\textwidth,angle=-90]{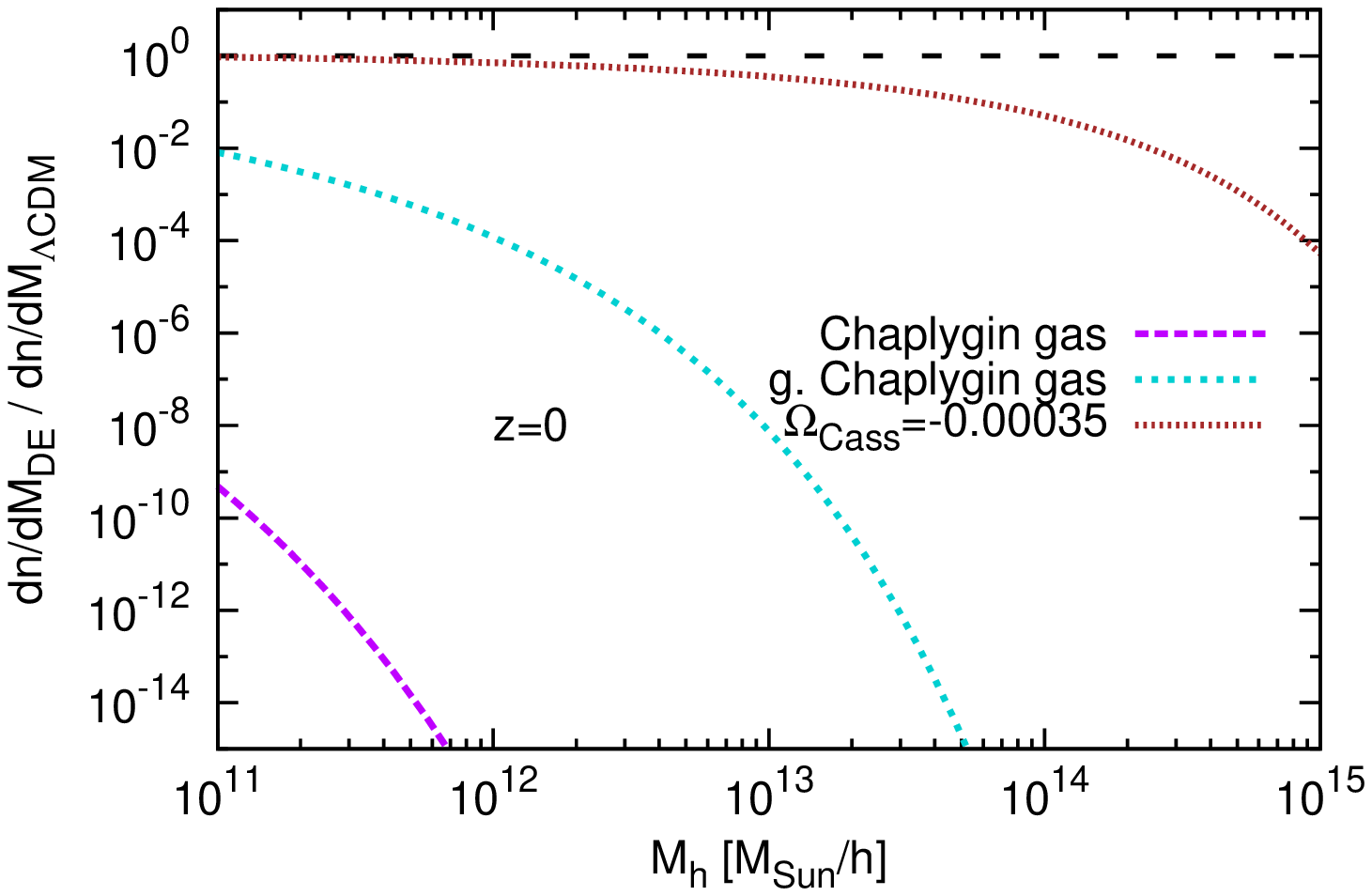} &
 \includegraphics[width=0.22\textwidth,angle=-90]{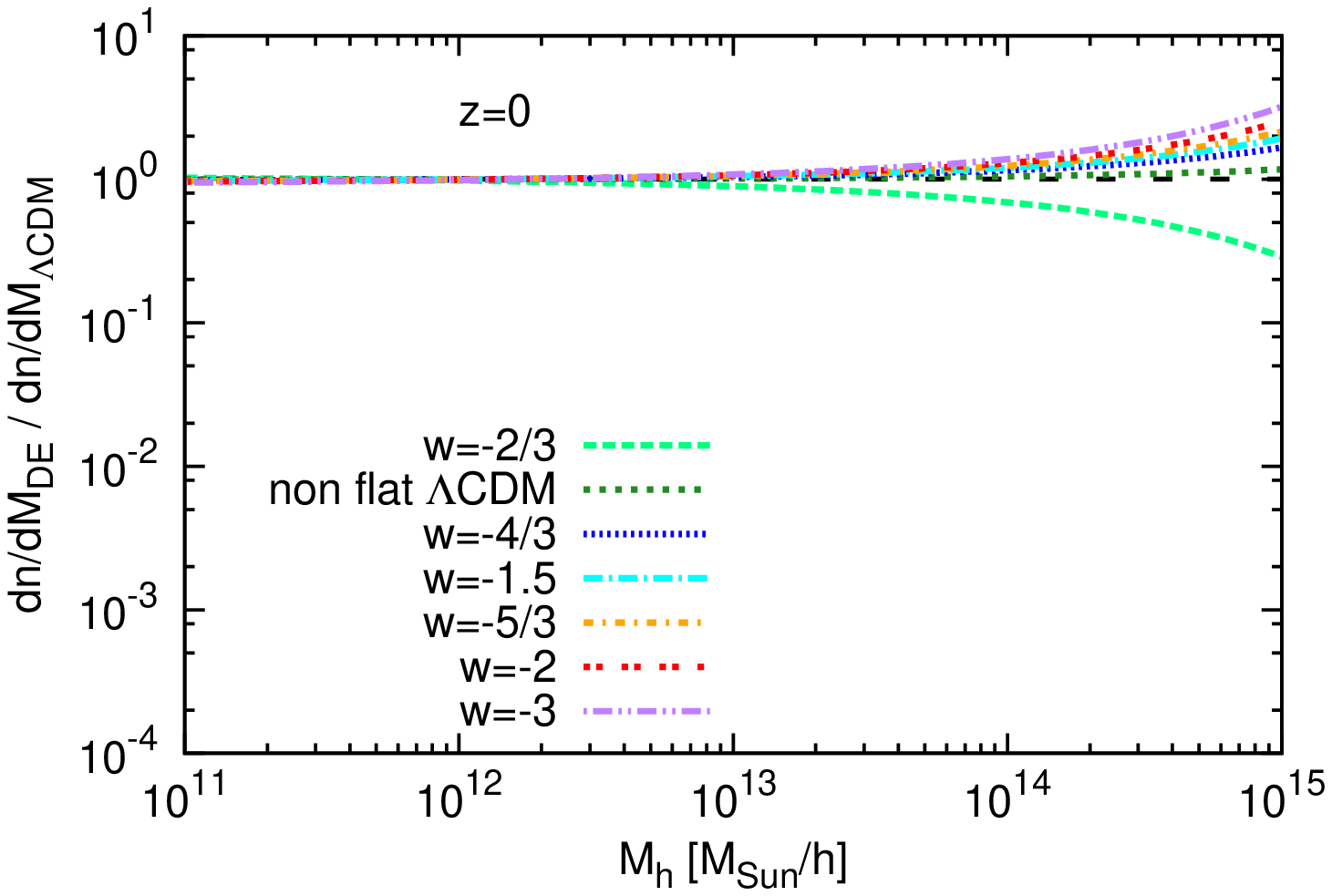}\\
 \includegraphics[width=0.22\textwidth,angle=-90]{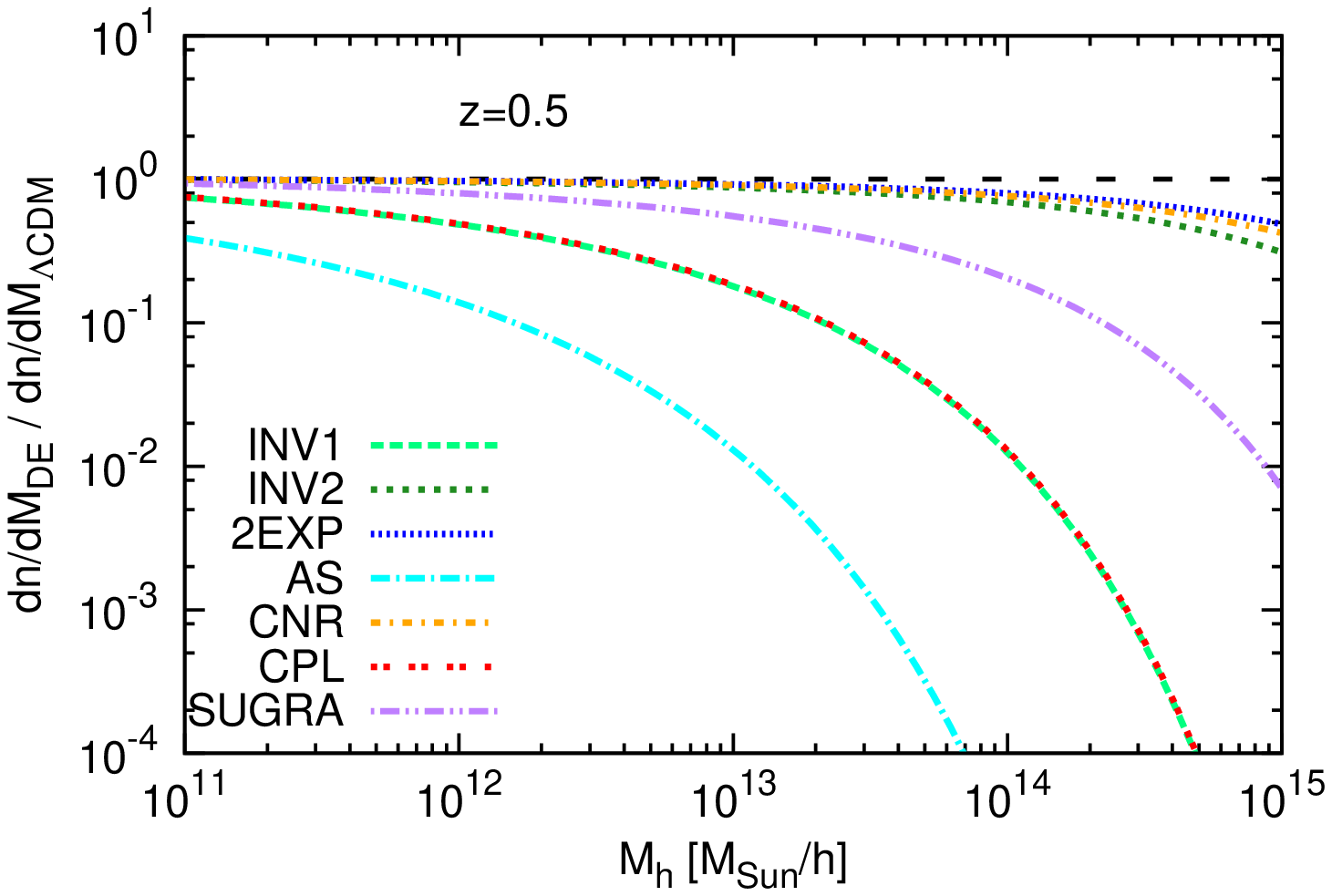} &
 \includegraphics[width=0.22\textwidth,angle=-90]{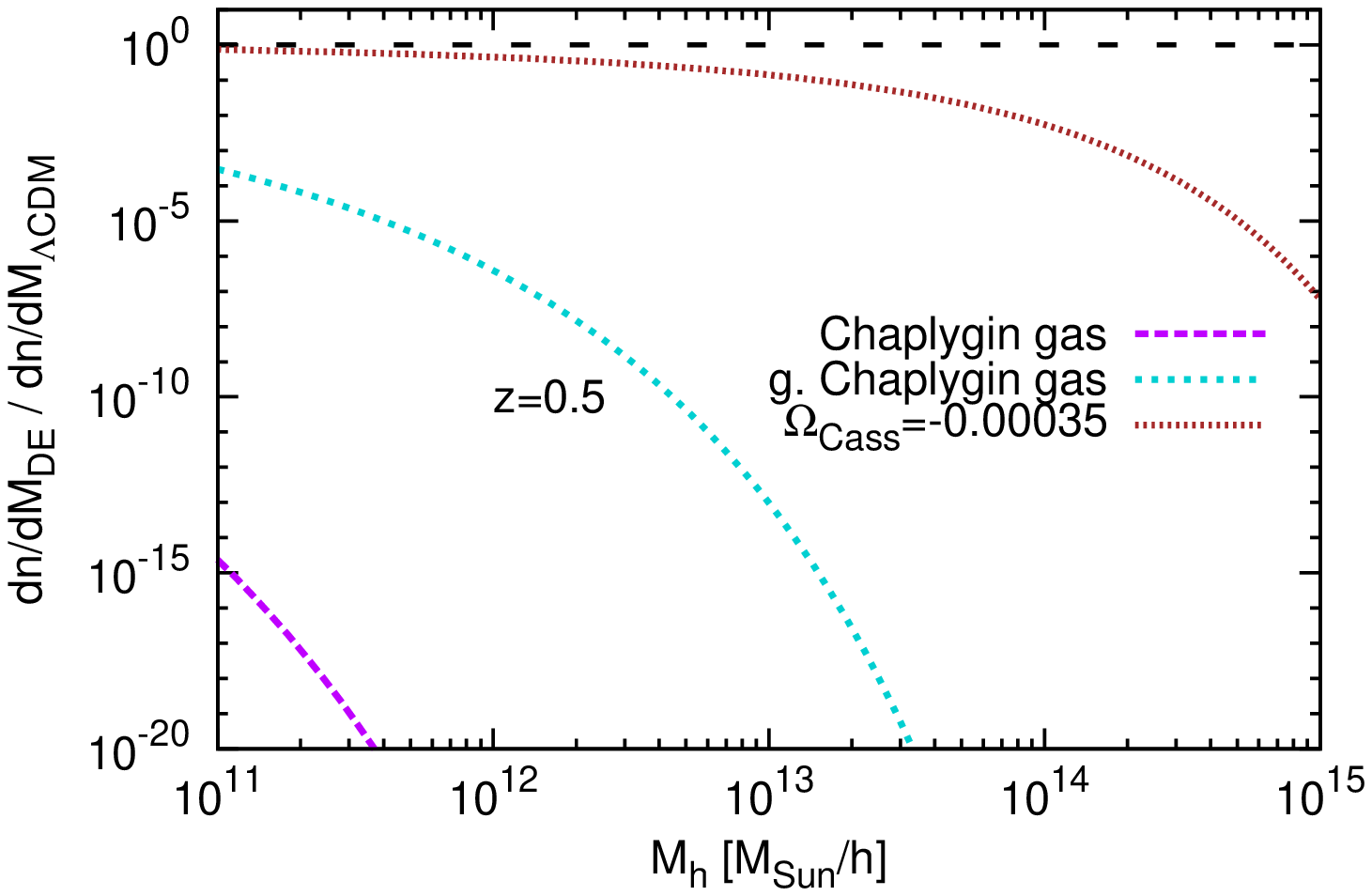} &
 \includegraphics[width=0.22\textwidth,angle=-90]{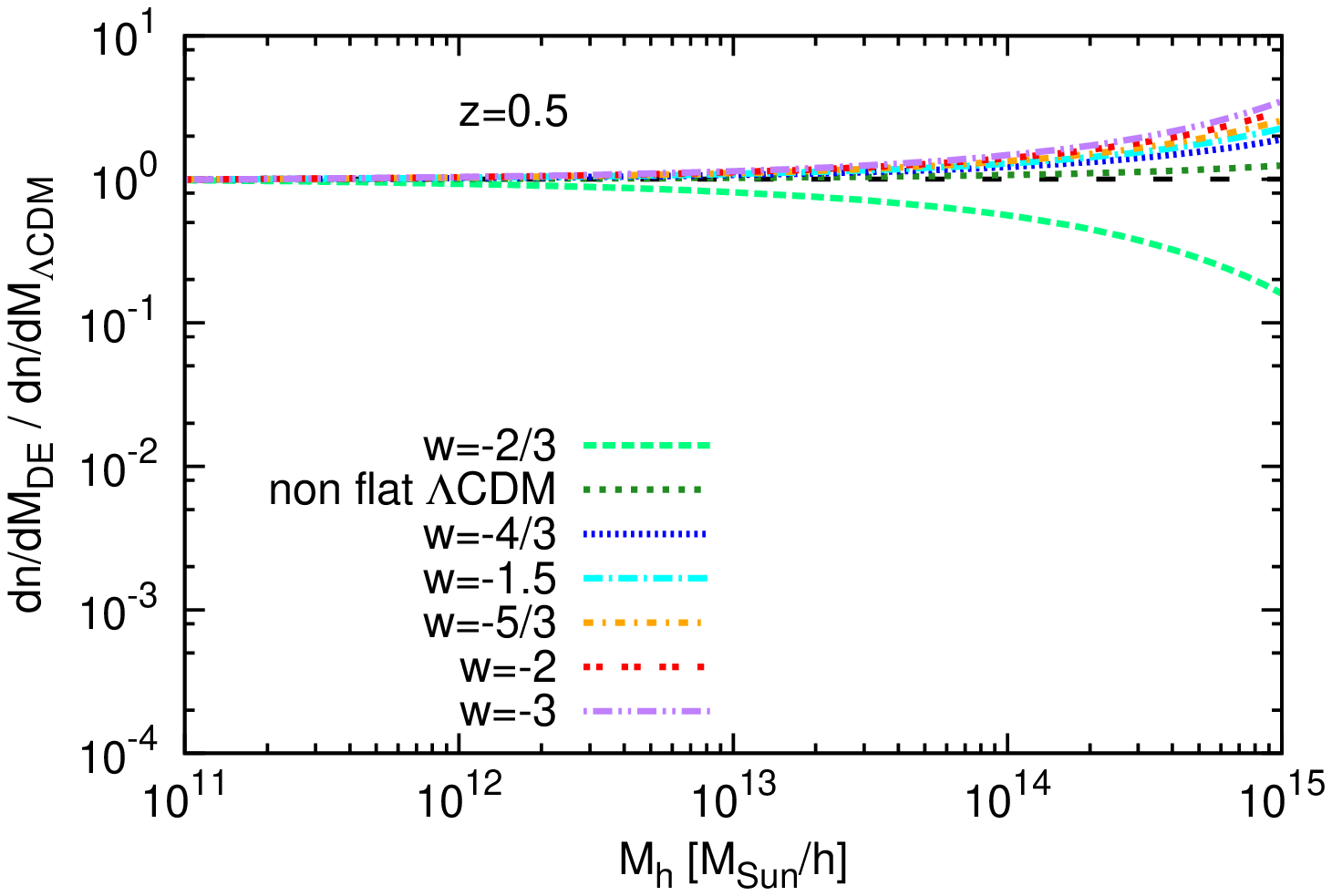}\\
 \includegraphics[width=0.22\textwidth,angle=-90]{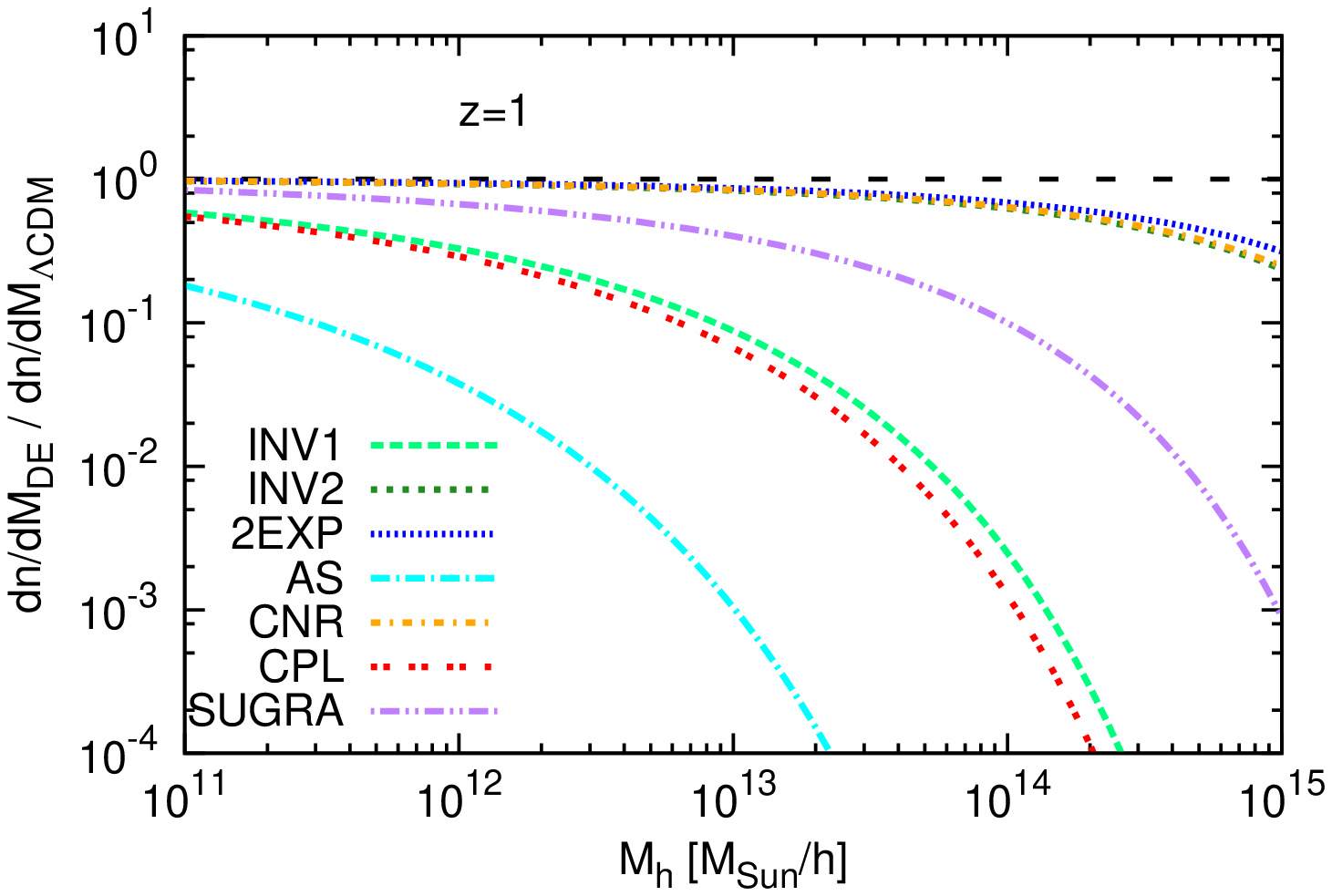} &
 \includegraphics[width=0.22\textwidth,angle=-90]{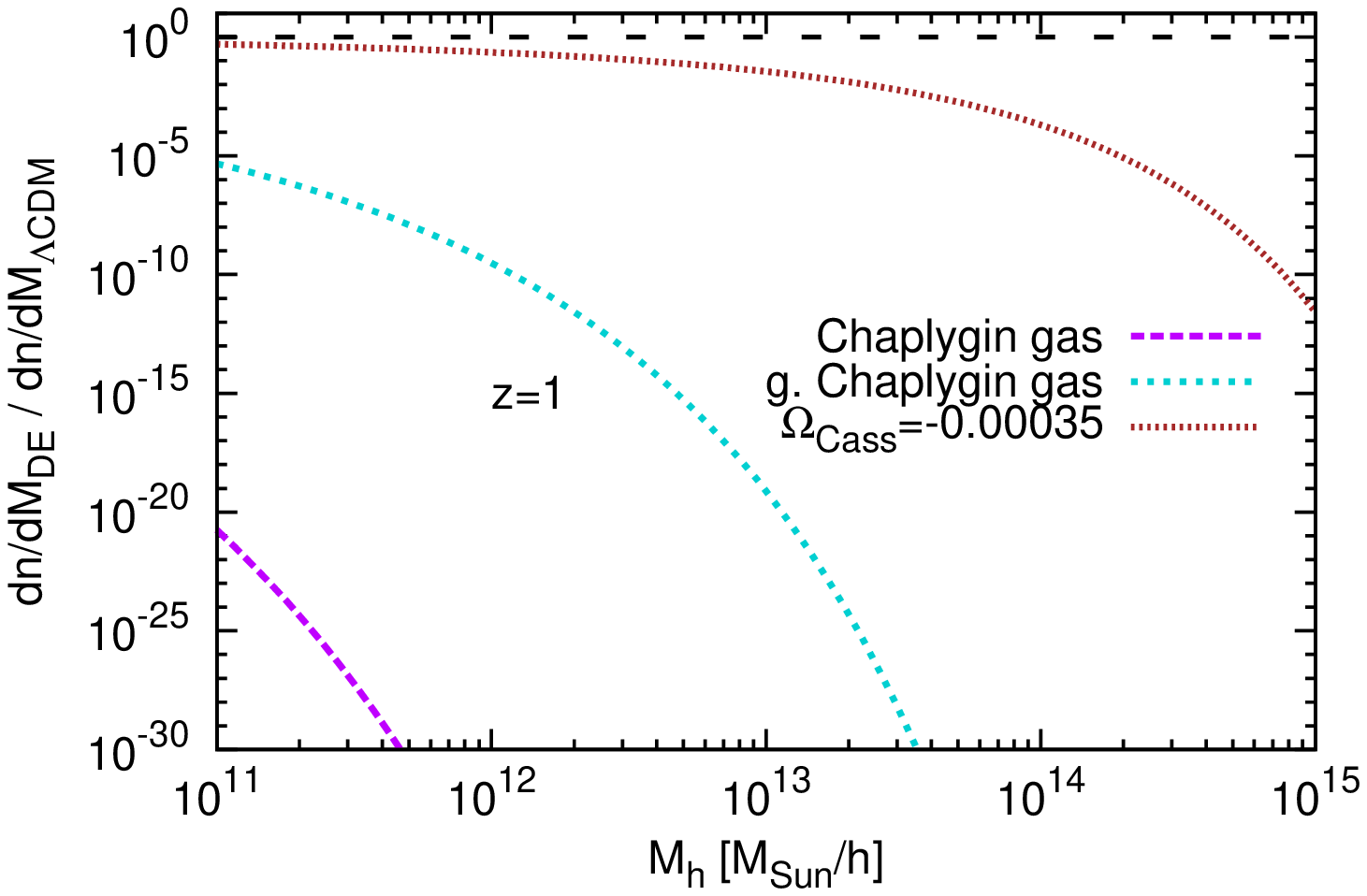} &
 \includegraphics[width=0.22\textwidth,angle=-90]{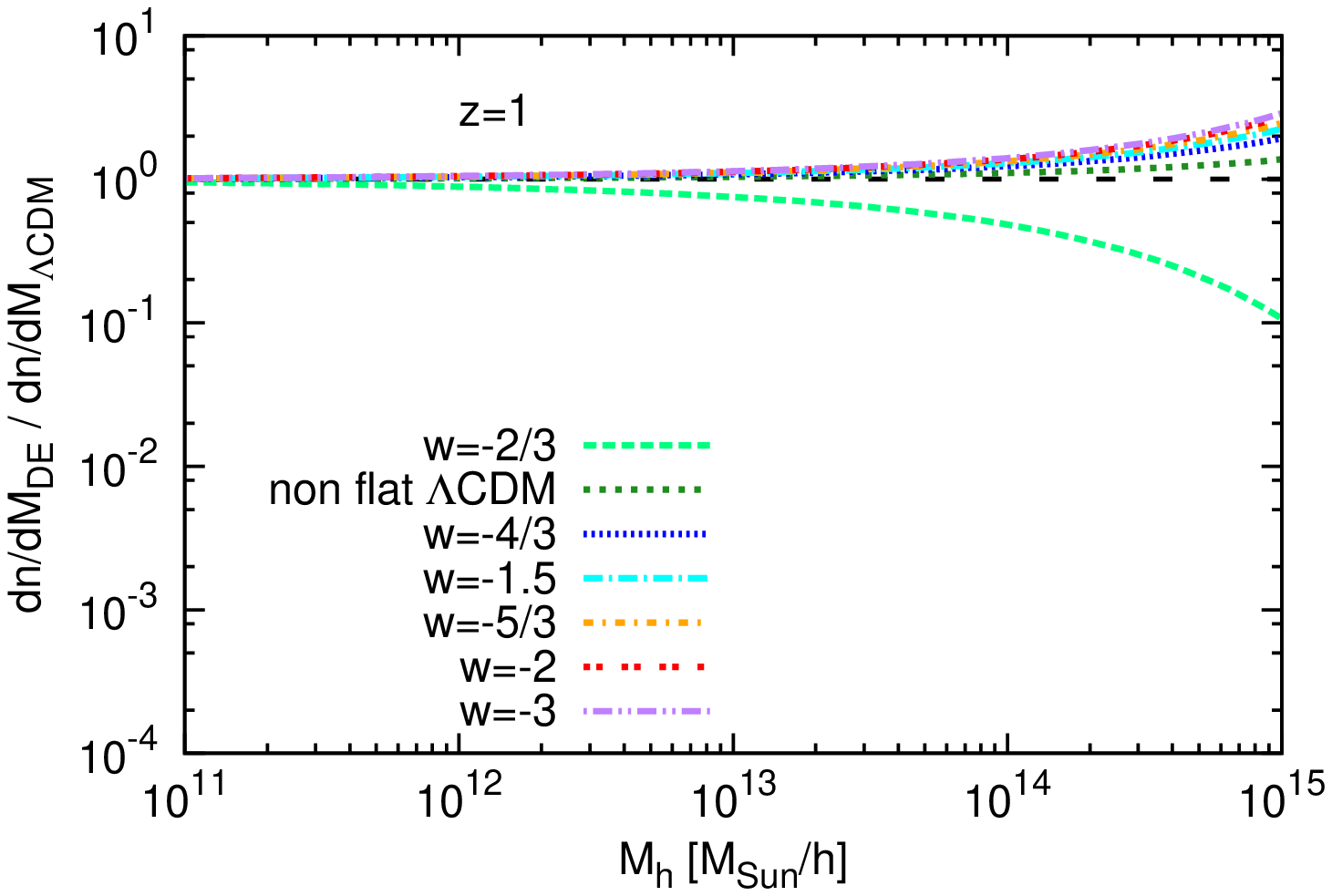}
 \end{array}$
 \caption{Differential mass function for the dark energy models considered in this work in the extended SCM for
different redshifts: $z=0$ (upper panels), $z=0.5$ (middle panels), $z=1$ (lower panels). Panels on the left (right)
show the quintessence (phantom and non-flat $\Lambda$CDM) models, the central panels show the models described by a
(generalised) Chaplygin gas and by the Casimir effect. We refer to Fig.~\ref{fig:spc} for line-styles and colours.}
 \label{fig:mfDE}
\end{figure*}

In Fig.~\ref{fig:mfDE} we show the ratios, at the three different redshifts considered, between the different dark
energy models here considered and the extended $\Lambda$CDM model, where the term $\sigma^2-\omega^2$ is included.

As we can notice, the (generalised) Chaplygin gas shows a huge suppression of structures at all redshifts, making
therefore this model ruled out in the extended SCM. All quintessence models have a lack of high mass objects. While
this is not severe at all for the INV2, 2EXP and CNR model, all the others have a suppression of several orders of
magnitude, increasing with the increase on the redshift. The most suppressed model is the AS model, that shows also the
most different $\delta_{\rm c}$ from the extended $\Lambda$CDM model.\\
Regarding the phantom models, differences are at most of a factor of 4-5. While the model with $w=-2/3$ shows a
decrease in structures, the phantom models show an increase. Differences are significant in general only for high masses
$M\gtrsim 10^{14}~M_{\odot}/h$ while for the model with $w=-2/3$, they are evident already at $M\approx 10^{13}$ for
$z=1$.\\
We also notice that a small amount of curvature has a very little effect on the number of objects, as differences are
of the order of few percent even for cluster scales.

Out results can be easily interpreted in terms of the different matter power spectrum normalizations. The Chaplygin gas
has an extremely low normalization ($\sigma_8=0.066$) making therefore very unlucky that structures could form in such
a universe. Phantom models instead show an higher normalization, making therefore easier to have high mass objects.
Moreover one has to take into account that now the linear overdensity parameter $\delta_{\rm c}$ is modified and very
strong differences will reflect in the differential mass function (see for example Fig.~\ref{fig:spc}).

A direct comparison between the results in this work and the ones described in \cite{Pace2010} cannot be made for
several reasons. In particular, the power spectra here were normalised in order to have the same amplitude of
fluctuations at the CMB epoch while in \cite{Pace2010} a different normalisation was adopted. There the power
spectrum was normalised in order to have nearly the same mass function at $z=0$. This implies that effects of
dark energy will be important only at high redshifts while in our case we see substantial differences already at low
redshifts as expected. In addition, here we just limit ourselves to the study of the differential mass function and we 
do not investigate the cumulative number of halos. This is because we do not want to have our results affected by volume
effects. Note also that due to the choice of normalization in \cite{Pace2010}, the quintessence models will predict more
objects than the $\Lambda$CDM at high redshifts (not shown here).

\section{Conclusions}\label{sect:conclusions}

In this work, we study the impact of the term $\sigma^2-\omega^2$ on the spherical collapse parameters, namely the
linear overdensity parameter $\delta_{\mathrm{c}}$ and the virial overdensity parameter $\Delta_\mathrm{V}$ and how this
reflects on the number of objects via the mass function formalism for a broad class of dark energy models, already
studied in the spherical limit, by \cite{Pace2010}.

We assume that only the dark matter component is clustering and that dark energy is only at the background level,
therefore affecting only the time evolution of the Universe. Doing this, we implicitly assumed that eventual
perturbations in the dark energy component can be neglected.

We showed that the non-linear term considered opposes to the collapse and this is reflected by higher values of the
linear overdensity parameter with respect to the spherical case. Modifications are quite substantial, of the order also
of 40\% for the $\Lambda$CDM model. In general the effect of dark energy is to lower the value of $\delta_{\mathrm{c}}$
with respect to the $\Lambda$CDM model and we see that this is also the case also in the extended SCM. Despite the
values of the linear overdensity parameter are higher now than in the non-rotating case, we see that in general dark
energy still lowers its value. This is the case only if $w>-1$, while for phantom models the super-negative equation of
state slows down the collapse.

{In order to appreciate the interplay between the different dark energy equation-of-state parameter and the power
spectrum normlization, it is interesting to compare models having approximately the same power spectrum normalization.
For example, this is the case for the models INV1, CPL and Casimir (see Table~\ref{tab:sigma} for the exact values).
Having very similar power spectra normalizations, from Fig.~\ref{fig:mfDE} we see that the ratio with the $\Lambda$CDM
mass function gives in general very similar results. Models INV1 and CPL are very similar to each other and in general
similar to the Casimir model. Differences between these models can be seen more clearly in the evolution of 
$\delta_{\rm c}$ in Fig.~\ref{fig:spc}. This shows that often differences in the models are hidden by the power
spectrum normalization. In general, taking into account this caveat, for other models differences in the
differential mass function are due to the influence of the dark energy component (see also comments at the end of
Sect.~\ref{sect:results}).}

As expected, such differences reflect in the number of objects. Since with respect to the standard case the only
quantity to be changed is $\delta_{\mathrm{c}}$, we can easily study the impact of a rotation term on structure
formation. We show this in Fig.~\ref{fig:mfDE}. The term $\sigma^2-\omega^2$ suppresses, as expected, the high mass tail
of the mass function, since rare events are more sensitive to the background cosmology and to the collapse process. In
general low masses objects are not severely affected by rotation, but a noteworthy counterexample is given by the
(generalized) Chaplygin gas and AS model where we observe a suppression in the number of objects already of several
orders of magnitude for galactic masses.

We conclude therefore that the term $\sigma^2-\omega^2$ has a strong impact on structure formation and that it is worth
to investigate different parametrizations for the additional term.

\section*{Acknowledgements}
The authors would like to thank the referee Y. Ascasibar for the valuable comments that improved our manuscript. ADP is 
partially supported by a visiting research fellowship from FAPESP (grant 2011/20688-1), and wishes also to thank 
the Astronomy Department of S\~ao Paulo University for the facilities and hospitality. FP is supported by STFC grant ST/
H002774/1, and JASL is also partially supported by CNPq and FAPESP under grants 304792/2003-9 and 04/13668-0. 

\bibliographystyle{mn2e}

\label{lastpage}

\end{document}